\documentclass[11pt]{article}
\usepackage{multirow}
\usepackage{graphicx}
\usepackage{tikz}
\usetikzlibrary{decorations.pathmorphing}
\usetikzlibrary{shapes.misc}
\usetikzlibrary{arrows}

\usepackage{floatrow}
\usepackage{multicol} 
\usepackage{color}
\usepackage{latexsym}

\definecolor{rossos}{cmyk}{0,1,1,0.55}
\definecolor{rossoc}{cmyk}{0,0.5,1,0.2}
\definecolor{blu}{cmyk}{1,1,0,0.3}
\definecolor{blus}{cmyk}{1,1,0,0.6}
\definecolor{blucc}{cmyk}{1,0.4,0.2,0}
\definecolor{viola}{cmyk}{0,1,0,0.6}
\definecolor{viola2}{cmyk}{0,1,0.2,0.6}
\definecolor{verde}{cmyk}{0.92,0,0.59,0.25}
\definecolor{verdec}{cmyk}{0.92,0,0.59,0.15}
\definecolor{verdes}{cmyk}{0.92,0,0.59,0.4}
\font\tenrsfs=rsfs10 at 12pt
\font\sevenrsfs=rsfs7
\font\fiversfs=rsfs5
\newfam\rsfsfam 
\textfont\rsfsfam=\tenrsfs
\scriptfont\rsfsfam=\sevenrsfs
\scriptscriptfont\rsfsfam=\fiversfs
\def\mathscr#1{{\fam\rsfsfam\relax#1}}

\def\circa#1{\,\raise.3ex\hbox{$#1$\kern-.75em\lower1ex\hbox{$\sim$}}\,}

\usepackage{amsmath,latexsym,amssymb,color,hyperref,graphicx}

\newcommand{\be}{\begin{equation}}
\newcommand{\ee}{\end{equation}}
\newcommand{\bea}{\begin{eqnarray}}
\newcommand{\ena}{\end{eqnarray}}
\newcommand{\no}{\noindent}

\renewcommand\o{\omega}  

\renewcommand\b{\beta}

\renewcommand\k{\ensuremath{\kappa}}

\newcommand\C{{\ensuremath{\cal C}}}

\newcommand{\de}{\partial}
\newcommand{\ha}{\frac{1}{2}}

\newcommand{\ba}{\begin{eqnarray}}
\newcommand{\ea}{\end{eqnarray}}

\newcommand{\vf}{\varphi}

\makeatletter
\def\ps@mine{%
    \def\@oddfoot{\hfil\thepage\hfil}\let\@evenfoot\@oddfoot
    \let\@oddhead\@evenhead%
    \let\@mkboth\@gobbletwo
    \let\sectionmark\@gobble
    \let\subsectionmark\@gobble
    }
\renewcommand\section{\@startsection {section}{1}{\z@}%
                                   {-3.5ex \@plus -1ex \@minus -.2ex}%
                                   {2ex \@plus.2ex}%
                                   {\normalfont\large\sffamily\bfseries}}
\renewcommand\subsection{\@startsection {subsection}{1}{\z@}%
                                   {-3.5ex \@plus -1ex \@minus -.2ex}%
                                   {2ex \@plus.2ex}%
                                   {\normalfont\sffamily\bfseries}}
                                   
                                   \renewcommand\subsubsection{\@startsection {subsubsection}{1}{\z@}%
                                   {-3.5ex \@plus -1ex \@minus -.2ex}%
                                   {2ex \@plus.2ex}%
                                   {\normalfont\sffamily\bfseries}}
\makeatother

\numberwithin{equation}{section}

\tikzset{
    photon/.style={decorate, decoration={snake}, draw=black} 
}
\tikzset{cross/.style={cross out, draw=black, minimum size=2*(#1-\pgflinewidth), inner sep=0pt, outer sep=0pt},
cross/.default={1pt}}

\usepackage{subcaption}
\begin{document}

\thispagestyle{empty}
\vspace*{-2.5cm}
\begin{minipage}{.45\linewidth}

\end{minipage}
\vspace{2.5cm}

\begin{center}
  {\huge\sffamily\bfseries Classical and Quantum Dynamics  of
    Gyroscopic Systems and Dark Energy}
\end{center}
 
 \vspace{0.5cm}
 
 \begin{center} 
 {\sffamily\bfseries \large  Denis Comelli}$^{a}$,
{\sffamily\bfseries \large  Maicol Di Giambattista}$^{b,c}$, 
 {\sffamily\bfseries \large  Luigi Pilo}$^{b,c}$,
 \\[2ex]
$^a$ INFN, Sezione di Ferrara, I-44122 Ferrara, Italy \\\vspace{0.3cm}
$^b$ Dipartimento di Scienze Fisiche e Chimiche, Universit\`a degli
Studi dell'Aquila,  Via Vetoio I-67100 L'Aquila, Italy\\\vspace{0.3cm}
$^c$ INFN, Laboratori Nazionali del Gran Sasso, Via G. Alcitelli 22 I-67100 Assergi, Italy\\\vspace{0.3cm}
 {\tt comelli@fe.infn.it}, {\tt maicol.digiambattista@aquila.infn.it}, {\tt luigi.pilo@aquila.infn.it},
\end{center}

\vspace{0.7cm}

\begin{center}
{\small \today}
\end{center}

\vspace{0.7cm}

\abstract{
\no Gyroscopic systems in classical and quantum field theory are characterized  by the presence of
  at least two scalar degrees of
freedom and  by terms that mix fields and
their time derivatives  in the quadratic Lagrangian.
In Minkowski spacetime,   they naturally appear
 in the presence of a coupling among fields with
time-dependent vacuum expectation values  and fields with
space-dependent vacuum expectation values, breaking spontaneously
Lorentz symmetry; this is the case for a  supersolid.
 In a cosmological background
a gyroscopic system can also arise from the time dependence of
non-diagonal kinetic and mass matrices. 
We study the classical and quantum dynamics computing the correlation
functions on the vacuum state that minimizes  the energy. 
Two 
regions of stability in parameter space are found: in one region, dubbed {\it
normal}, the
Hamiltonian is positive defined,  while in the second region, dubbed {\it
anomalous}, it has no definite sign. Interestingly, in the anomalous
region the 2-point correlation function exhibits a resonant behaviour
in a certain region of parameter space.
 We show that   as dynamical a dark energy  (with an exact equation of state $w=-1$)
arises naturally   as a gyroscopic system.

}
\clearpage

\section{Introduction}
The study of quadratic Lagrangian systems is the starting point for the analysis  of  classical/quantum solvable systems.
Gyroscopic systems are a particular class of Lagrangians/Hamiltonians, characterized  by rather surprising features even at linear level dynamics. 
 The above systems~\cite{Lancaster} (defined on a Minkowski
 background) are characterized  by $N$-degrees of freedom with $N>1$
 and their dynamics is described by the  following Lagrangian and  corresponding equations of motion
 \be
{\cal L}=\frac{1}{2} \dot \varphi^t\,{\cal K}\,\dot \varphi+\varphi^t\,{\cal D}\,\dot \varphi-\frac{1}{2}\,\varphi^t\,{\cal M}\, \varphi
\quad \to \quad{\cal K}\,\ddot{\varphi} -2 \, {\cal D} \,
\dot{\varphi} + {\cal M} \, \varphi =0 \, , \qquad \varphi^t
=(\varphi_1, \cdots, \varphi_N ) \, ;
\ee
where ${\cal K} $, ${\cal D} $ and ${\cal M} $ are
  $N \times N$ constant matrices with the following properties:
  ${\cal K} $ is symmetric and positive defined,
${\cal D} $ is antisymmetric while ${\cal M} $ is just
symmetric. 
\\
The defining property is the
presence of non-dissipative  velocity-dependent forces, described, in the
classical mechanics picture, by the antisymmetric matrix $ {\cal D}\neq 0$. 
Gyroscopic systems can be found in many area of physics and
engineering\footnote{Dissipation can be described by adding to ${\cal D}$ a
  symmetric part.}, for a fascinating review see~\cite{Krechetnikov:2007zz}.
 Such systems have a number of unusual features.
\begin{itemize}
 \item  Stability is realized in a peculiar way.
 Beyond  a normal stability region where the mass matrix ${\cal M}$ is positive defined 
 (we call such a parameter space region, the {\it normal stability
   region}), it exists a stability region also for negative defined
 mass matrix ${\cal M}$
 and the matrix ${\cal D}$ results larger that a critical value ${\cal
   D}_c$ (${\cal D}\geq {\cal D}_c$); we call such a parameter space
 region, the {\it gyroscopic or anomalous region}.
 \item Despite of the fact the Hamiltonian is time-independent and
   then system is  conservative,  time reversal symmetry is violated
   by the presence of single time derivative operator  $\vf\,\dot\vf$
   proportional to the ${\cal D}$ matrix~\cite{Roberts:2016ofs} in the
   Lagrangian.
\item 
  On a Minkowski background, as a consequence of the spontaneous
  breaking of Lorentz invariance down to the rotational group, the elementary excitations can be
  interpreted as phonon-like modes of a supersolid; namely a solid
  coupled with a superfluid. Symmetry arguments imply the term ${\cal
    D}$ in the quadratic Lagrangian exists
only when scalar fields  with non-trivial vacuum configurations are coupled together.
\item
  The quantisation of the system with a Fock representation of the
  canonical commutation relation is feasible only after a suitable 
  diagonalisation of the Hamiltonian which can be written 
as a set of decoupled harmonic oscillators. In the Lagrangian approach
such a decoupling cannot be achieved.
 As a general result,  while in the  {\it normal stability region} the
 Hamiltonian is positive defined, in the  {\it gyroscopic (or
   anomalous) region} the Hamiltonian results negative defined showing
 an intriguing connection of a gyroscopic system with the
 Pais-Uhlenbeck oscillator. Interestingly, the long standing problem
 of the stability for interaction  Pais-Uhlenbeck oscillator was
 recently reconsidered~\cite{Gross:2020tph} and the resonant behavior
 of the 2-point correlation that we found in the anomalous region
 plays an important role.
\end{itemize}
 On a generic time-dependent background the matrices ${\cal K},\;{\cal
   D}$ and ${\cal M}$ are naturally time-dependent and
 the definition of a gyroscopic system is 
 ambiguous due to the possibility of  performing 
 time- dependent field transformations.
Focusing  on the case of  two scalar degrees
 of freedom, which represent  the minimal field content for a
 gyroscopic system\footnote{The result be generalised to the case a
 generic even number scalar fields.}, we show that it possible by
   a suitable  set of time-dependent Lagrangian field transformation to
  bring the system in a  {\it canonical} form where 
  the kinetic matrix ${\cal K}$ is the identity, the mass matrix is
  diagonal and ${\cal D}_c= d_c \, \epsilon_{ab}$ where $\epsilon_{ab}=-\epsilon_{ba},\;a,b=1,2$  and $\epsilon_{12}=1$.
 In its canonical form, a system is unambiguously gyroscopic if $d_c \neq
0$ and it is  characterized by three time dependent parameters: $d_c$ and the
diagonal entries of the mass matrix.
In this framework we study when  $d_c \neq
0$.
Generally speaking,  gyroscopic systems can manifest when some
of the fields acquire a non-trivial
background spacetime dependent ``vacuum'', a behavior present in many multi-field systems, see for example the effective description of
media~\cite{Matarrese:1984zw,Son:2000ht, Dubovsky:2011sj,Andersson:2020phh}, massive gravity~\cite{Dubovsky:2005xd,ussgf,Celoria:2017bbh}
single~\cite{Cheung:2007st} and  multi-field
inflation~\cite{Senatore:2010wk}, solid and supersolid
inflation~\cite{Endlich:2012pz,Celoria:2021cxq,Celoria:2020diz} and holography~\cite{Baggioli:2019abx,Baggioli:2022aft}.

\section{Scalar Fields and Vacuum  Configurations}
\label{origin}
We are interested in the study of systems characterized  by a set of  $N$ scalar fields $\{ \Phi^A , \;
A=1, \cdots , N\}$ such that some of them acquire a non-trivial
spacetime dependent vacuum expectation value (vev) that describes the background
configuration of the system.
Therefore the fields are split
in a background configuration $\phi^A$ plus a fluctuation
$\varphi^A$
\be
\Phi^A = \phi^A + \varphi^A \, .
\ee
The fields $\{\varphi^A \}$ are associated   to 
the classical/quantum small fluctuations.
For instance, in the effective description of fluid
dynamics, the background configuration of the fluid
is described by  $\phi^A$  while the phonon excitations are described
by  $\varphi^A$. It is useful to distinguish  the fields according to
the nature of their vev; namely
\begin{itemize}
\item fields with zero vev  $\phi^A=0$  and fluctuations $\varphi^A \equiv Z^A $;
\item fields with a time-dependent vev, for example  $\phi^A=c^A\, t$,   and
  fluctuations  $\varphi^A\equiv T^A$;
\item fields with $\vec x$-dependent vev $\phi^A=c^A_j \, x^j $ and  fluctuations  $ \varphi^A\equiv S^A$.
\end{itemize}
Our goal is to study the dynamics of $\varphi^A$.
We shall consider the case   where the underlying symmetry group of
the background spacetime is partial broken by scalar fields
configuration  $\phi^A$ in such a way that spatial translations and rotations stay
unbroken.
Barring accidental cancellations, translational invariance requires
that spatial derivatives $\de_i \phi^A$ must be constant. 
In the case of fields with ${\vec x}$-dependent vev it implies that $ \phi^A\equiv \phi_n^i=x^i$. In order to
automatically implement such a constraint, we always
require a shift symmetry for the fields with an ${\vec x}$-spatial vev, namely
\be
{\rm Spatial\, vev\, fields:}\quad\Phi^i_n \to \Phi_n^i + \text{constant} \qquad \text{if }\quad
\phi_n^i=x^i \, , \qquad i=1,2,3
\label{spshift}
\ee
where now $n$ denote the number of different fields.
   In other words, being   interested in systems where spatial rotations are always
unbroken, in 3+1 dimensions the minimal number of ${\vec x}$-spatial vev fields  is three. A triplet of scalar
fields $\{ \Phi^a, \; a=1,2,3 \}$ is transforming as the fundamental  representation of
an internal $SO(3)_I$ symmetry group of the system. The vev
induces the following spontaneous breaking pattern
\be
SO(3) \times
SO(3)_I \to SO(3)_D \, .
\label{bp}
\ee
 In general the total
number of fields  with an $\vec x$-spatial vev consists of $n$ triplet of
$SO(3)_I$. 
Then for each fluctuation  we can use the  Helmholtz decomposition
\be\label{Helmholtz}
\varphi^A\equiv\varphi^{i}_n=S^i_n\equiv \frac{\partial_{i}}{\sqrt{\vec\nabla^2}}\,S_n+{V}_n^i,\qquad \partial_i{V}_n^i=0, \qquad i=1,2,3
\ee to extract the scalar $S_n$ and the vector ${V}_n^i$ components.
Being interested   in the scalar sector, only $S_n$ will be relevant for us.
\\
For the fields with a time dependent vev,  a shift symmetry is not strictly necessary, in particular when also
the background is breaking time diff as in FRW. On the contrary when 
we work in a Minkowski spacetime and as soon as we require  an EFT
with thermodynamical properties \cite{Dubovsky:2011sj}, also the
temporal  vev fields have to be shift symmetric.\\
The dynamics of the
fluctuations can be found by studying the  structure of all operators (with up to two derivatives)  consistent with rotations. 
At the quadratic level we have the following  classification scheme:
\begin{itemize}
\item Operators with no derivatives: 
\be
{\cal O}_0^{AB}=\varphi^A\,\varphi^B \, .
\ee 
\item Operators with one derivative: 
\be
{\cal O}_x^{AB}= \de_i \varphi^A\, \varphi^B \, ,  \qquad  {\cal O}_t^{AB} =
\varphi^A\,\dot{\varphi}^B   \, .
\ee
\item Operators with two derivatives: 
\be
{\cal O}_{xx}^{AB} =
\de_i \varphi^A \,\de_j \varphi^B \, , \qquad
 {\cal O}_{xt}^{AB} 
=\de_i \varphi^A  \, \dot{\varphi}^B \,  , \qquad 
{\cal O}_{tt}^{AB}  = \dot{\varphi}^A  \dot{\varphi}^B \, .
 \ee
\end{itemize}
To produce a rotational invariant quadratic Lagrangian ${\cal L}$,
the indices in $\{ {\cal O}_n \}$ should be saturated by using the
only two available invariant tensors $\delta_{ij}$ and
$\epsilon_{ijk}$. The result is 
\be
{\cal L}= {\cal L}_2 + {\cal L}_1-{\cal L}_0  \, ;
\ee
where
\bea
&& {\cal L}_2 = {\cal K}\cdot {\cal O}_{tt}  \, ;\\[.2cm]
&& {\cal L}_1= {\cal D}^{(t)}\cdot {\cal O}_{t} +{\cal D}^{(tx)}\cdot
{\cal O}_{xt} \, ;\\[.2cm]
&&{\cal L}_0=  {\cal M}^{(0)}\cdot {\cal O}_0 +
 {\cal M}^{(x)}\cdot {\cal O}_{x} +
 {\cal M}^{(xx)}\cdot {\cal O}_{xx} \, ;
\ea
and $\cdot$ stands for rotational invariant contractions of the
relevant indices. The term $ {\cal L}_2$ has two time derivatives and
represents the kinetic term while  $ {\cal L}_0$ with no time
derivatives is a mass term. Beside a rather standard kinetic matrix
${\cal K}_{AB}$ and a mass matrix ${\cal M}_{AB}$, the peculiar term
is the one linear in  the time derivative of the fluctuations 
and proportional to ${\cal D}_{AB}$; the presence of such a term is the
defining property of a gyroscopic system. Notice that the kinetic
matrix ${\cal K} $ and the mass matrix ${\cal M} $ are
symmetric by construction, while one can take ${\cal D}=-{\cal D}^t $ by adding/subtracting a total derivative term.
Indeed, by splitting ${\cal D}$ as a symmetric ${\cal D}^{(S)}$ and an
antisymmetric ${\cal D}^{(A)}$ part, the former can be cast into a
mass term up to a total derivative 
\be
{\cal D}^{(S)}_{AB} \;\dot{\varphi}_A \;\varphi_B =\ha  {\cal
  D}^{(S)}_{AB}\; \frac{d}{d t} \left ( \varphi_A \varphi_B \right)=\ha
\frac{d}{d t} \left ( {\cal D}^{(S)}_{AB}\;  \varphi_A\, \varphi_B\right) -
\ha \dot{ {\cal D}}^{(S)}_{AB}\; \varphi_A\, \varphi_B \, .
\label{ibp}
\ee
When  the structure of the Lagrangian is further restricted by imposing
a shift symmetry on {\it all} scalar fields, namely
\be
\Phi^A \to \Phi^A + \text{constant},
\label{cshift}
\ee
then all operators with a  single or zero derivatives (temporal or spatial) are forbidden and the structure of the shift symmetric Lagrangian reduces to
\be
{\cal L}^{\text{shift}}= {\cal K}\cdot {\cal O}_{tt}  +
  {\cal D}^{(tx)}\cdot {\cal O}_{tx}  - 
 {\cal M}^{(xx)}\cdot {\cal O}_{xx}  \, .
 \ee
Notice that even if (\ref{cshift}) is imposed, the presence
of ${\cal D}^{(xt)}$ breaks time reversal symmetry.
In table \ref{T1} we show the structure of the quadratic Lagrangian
depending on the type of vevs and on the internal shift symmetries
imposed. In the general case the Lagrangian ${\cal L}$ contains
all the operators. 
In the Lorentz invariant (LI) case, when the very same
symmetries of Minkowski space are imposed, the system cannot be
gyroscopic. 
\begin{table}
\begin{tabular}{|c||cccccc|}
\hline
&${\cal K}$&${\cal D}^{(t)}$&${\cal D}^{(tx)}$ &${\cal
                                                 M}^{(0)}$&${\cal
                                                            M}^{(x)}$&${\cal
                                                                       M}^{(xx)}$\\
  \hline \hline
${\cal L} $&  \checkmark &  \checkmark&  \checkmark  &\checkmark
                                                                &
                                                                  \checkmark
                                                                                 &
                                                                                   \checkmark
  \\
\hline 
${\cal L}^{\text{shift}}$ &\checkmark & &  \checkmark  & & &  \checkmark\\\hline\hline
  LI&  \checkmark & & & & &  \checkmark\\
  \hline\hline
${\cal L }_{T}$&  \checkmark & \checkmark & &  \checkmark& &
                                                              \checkmark\\
  \hline
  ${\cal L}_{T}^{\text{shift}}$&  \checkmark & &  & & &  \checkmark\\
  \hline\hline
  ${\cal L}_{S}$&  \checkmark & & &  \checkmark& & \checkmark\\
  \hline
  ${\cal L}_{S}^{\text{shift}}$&  \checkmark & & & & & \checkmark\\
  \hline
\hline
${\cal L}_{TS}$&  \checkmark & & \checkmark & &  \checkmark&
                                                             \checkmark\\
  \hline
${\cal L}_{TS}^{\text{shift}}$&  \checkmark & & \checkmark & & & \checkmark \\
  \hline
\end{tabular}
\caption{Structure of the quadratic Lagrangian. The suffix {\it shift}
  indicates the presence of the complete shift symmetry (\ref{cshift}).
$T(S)$ stands for field with only time-dependent (space-dependent)
vev while $TS$ underline that fields with both time-dependent
and space-dependent vev are present. In the Lorentz invariant (LI) case the very same
symmetries of Minkowski space are imposed.}
\label{T1}
\end{table}

The special  cases with only $t$-dependent vev fields $T^A$ (case ${\cal L}_T$),
  $\vec x$-dependent vev fields $S_n$ (case ${\cal L}_S$) or both (case ${\cal L}_{TS}$) is presented.
For the rest of the paper we will assume the presence of two scalar degrees of freedom ($N=2$) so that all matrices will be 2$\times$2.
In Fourier space where
\be
\varphi(t,\pmb{x})=\int \frac{d^3k}{ (2\,\pi)^{3/2}}\;\varphi_{\pmb{ k}}(t )\;e^{i\,\pmb{k} \cdot \pmb{x}} \;, \qquad \qquad
\varphi =\begin{pmatrix} \varphi_1 \\ \varphi_2 \end{pmatrix} \,
\ee
the reality of the fields $\varphi(t,\pmb{x})=\varphi(t,\pmb{x})^*$ imposes that $\varphi_{\pmb{ k}}(t )=\varphi_{-\pmb{ k}}(t)^*$
or $\varphi_{-\pmb{ k}}(t )=\varphi_{\pmb{ k}}(t)^*$.
The Lagrangian (where only the time derivatives of the fields appear) takes the form (see
\ref{bexp} for more details) 
\be
{\cal L} = \ha \, \dot{\varphi}_{\pmb{ k}}^\dagger \, {\cal K} \, \dot{\varphi}_{\pmb{ k}}+ \varphi^\dagger_{\pmb{ k}}
\, {\cal D} \, \dot{\varphi}_{\pmb{ k}} - \ha \, \varphi^\dagger_{\pmb{ k}}
\, {\cal M} \, \varphi_{\pmb{ k}} \, ;
\label{quadlc}
\ee
where the ${\cal K}$, ${\cal D}$ and ${\cal M}$ matrices absorbed  the   ${\pmb{ k}}$ dependence of the spatial derivative of the fields. Further, the structure of the ${\cal D}$ matrix is fixed to be
\be
{\cal D}=d\;{\cal J},\qquad { \cal J}= \begin{pmatrix}
0&1\\
-1&0
\end{pmatrix} \, ;
\ee
and $d$ is a real parameter  (where the integration by parts (\ref{ibp})
has been used to have ${\cal D}$ antisymmetric).
Few comments are in order. Spatial derivatives $\partial_x$ in Fourier space
are replaced by $ i \, k$ and 
the ``mass'' matrix ${\cal M}$, kinetic mixing term ${\cal D}$ and the kinetic
matrix ${\cal K}$ are both time and $k$-dependent. 
 In
the following we will study the $k$ dependence of (\ref{quadlc}) in detail.
To end up this chapter we see that a gyroscopic system with ${\cal D}\neq0$ can be generated essentially
\begin{itemize}
\item when ${\cal D}^{(t)}\neq0$: in this case we need interactions between $T^A$ fields that are not protected by shift symmetry (typically present in FRW backgrounds and not in Minkowski space);
\item when ${\cal D}^{(tx)}\neq0$: in this case we need interactions between $T^A$ and $S_n$ fields and
are present also for shift symmetry Lagrangians (i.e. in Minkowski background and for the description of the Goldstone modes of ideal thermodynamical system).
\end{itemize}
This definition of gyroscopic systems is then valid both  in Minkowski space and in FRW backgrounds.

\section{Canonical form of a Gyroscopic system }
\label{sec-canform}
Taking into account the presence of non-trivial vacuum configuration,
one finds  that the most general  structure for the
quadratic Lagrangian   is of the form (\ref{quadlc}). We
  stress again that  to get   the form  (\ref{quadlc}) we have used
  only integration by parts without modifying the original equations  of motion.
  Consider now  a  general linear  Lagrangian field transformation 
  $\varphi\to F(t)\, \varphi$, with $F$ an invertible time-dependent matrix.
  Obviously, a  time-dependent transformation will lead again to a
  Lagrangian of the form  (\ref{quadlc}) but with different
  matrices $\tilde{\cal K}$, $\tilde{\cal D}$ and $\tilde{\cal M}$; in particular such a
  transformation can induce an effective $\tilde{\cal D}$ that
  characterises a gyroscopic system, even if it was zero in the original
  field variables. 
Indeed, the effect of the above transformation on the equations of
motion is the following
\be
\varphi= F(t)\,Q\quad \Rightarrow \quad ({\cal K}\;F)\;\ddot
Q-2\,\underbrace{({\cal D}\,F-{\cal K}\;\dot F)}_{\tilde{\cal
    D} }\,\dot Q+({\cal M}\,F+{\cal K}\,\ddot F+{\cal D}\,\dot
F)\,Q=0 \, .
\ee
When  $\dot F\neq0$ an effective $\tilde{\cal D} $ generically arises.
 As shown in appendix \ref{canform}, in order to remove such ambiguity
 one can always make 
a suitable linear Lagrangian field redefinition (time-dependent in
general) to put the
matrices entering (\ref{quadlc}) in the following {\it canonical form}
\be
{\cal K}\to \pmb{I}, \qquad {\cal D}\to { D} =d_c \, {\cal J},
\qquad {\cal M}\to { M} =\begin{pmatrix} m_1^2&0 \\ 0
  &m_2^2 \end{pmatrix} \, ;
\label{gyrcan}
\ee
where $\pmb{I}$ is the identity matrix. The argument can be
generalised to the case of $N\geq2$ degrees of freedom. Once the matrices in
(\ref{quadlc})  are in their canonical form (\ref{gyrcan}), we 
define a system ``gyroscopic'' if $ D  \neq 0$. 
A non-vanishing $
{ D}  \neq 0$ can originate
from the  ``canonisation'' of (\ref{quadlc}); in particular a
non-trivial $d_c$ is generated by a time-dependent diagonalisation of the original
kinetic and mass  matrices ${\cal K} $,  $ {\cal M}$ with
time-dependent rotation angles  $\theta_{\cal K}$ and
$\theta_{\cal M}$ respectively (see appendix \ref{canform} for details); namely
\be\label{dc}
d_c=\frac{d}{\det({\cal K})^{1/2}}-\frac{\text{Tr}({\cal
    K})}{\det({\cal K})^{1/2}} \, \dot\theta_{\cal
  K}-2\,\dot\theta_{\cal M}^2 \, .
\ee
Thus, a system   is  gyroscopic when at least one of the following
conditions are satisfied:
\begin{itemize}
\item the original ${\cal D} \neq 0$;
\item a non-trivial time dependence of the kinetic matrix such that
  $\dot\theta_{\cal K}\neq 0$;
\item a non-trivial time dependence of the mass matrix such that
  $\dot\theta_{\cal M} \neq 0$.
\end{itemize}
In a cosmological background, a typical situation where  $\{
\dot\theta_i \} $ are zero is 
when the matrices ${\cal K}$ or ${\cal M}$   are
\footnote{The mixing angle $\theta_\lambda$ for a generic $2\times 2$
  matrix $\lambda$,  is given by $\tan (2 \,
  \theta_\lambda)=\frac{2\,\lambda_{12}}{\lambda_{22}-\lambda_{11}}$
  with its time derivative
given by
$\dot\theta_\lambda=\frac{(\lambda_{22}-\lambda_{11})\;\dot\lambda_{12}-\lambda_{12}\;(\dot\lambda_{22}-\dot\lambda_{11})}{4\;\lambda_{12}^2+(\lambda_{22}-\lambda_{11})^2}$. It
is easy to identify when $\dot\theta_\lambda=0$.}: diagonal, there is an
overall time dependence, the diagonal entries are equal, the difference between the two diagonal elements is proportional to the off-diagonal one. 
The cases where the matrices are time-independent (and thus
$\dot\theta_i=0$) include Minkowski space  background and,  more in
general, a regime
where momenta are much higher than the inverse of the curvature scale.
This is the case for the definition of the Bunch-Davies vacuum in a de Sitter background (see chapter \ref{BDV}).
Unless explicitly stated we will consider a
gyroscopic system in the canonical form (\ref{gyrcan}).
One might think that after all the system (\ref{quadlc}) is rather
simple. However, as pointed out humorously by
Coleman~\cite{Coleman:2018mew}, quantum field theory is based on
different variations of the harmonic oscillator. The first evidence
that (\ref{quadlc}) is less trivial than it looks is that, as shown
in appendix \ref{tdeptr}, there is no
Lagrangian field redefinition to set ${\cal D}=0$ and at the same time having  a diagonal mass matrix; indeed, in a Minkowski background, setting ${\cal D}=0$ through a further time dependent linear transformation of the canonical fields generates a mass term with periodic time dependence (Floquet system).
While to find classical solutions is not
a problem, the 
quantisation is not straightforward.

\section{Stability of a time-independent Gyroscopic System}
\begin{figure}[htbp]
\centering
\includegraphics[width=0.4\columnwidth]{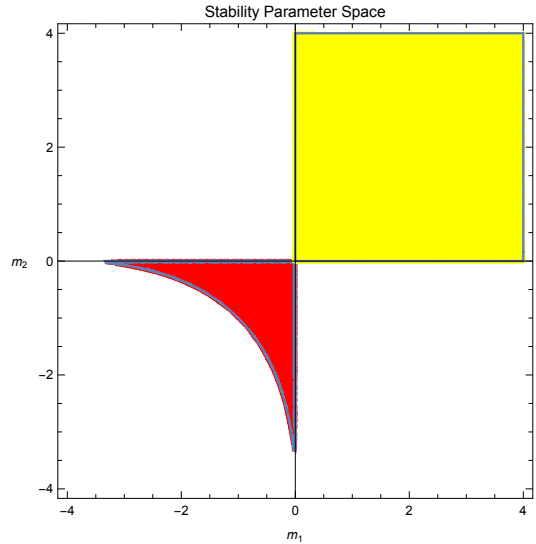}
\caption{For $d=1$, in yellow the parameter space corresponding to
  (\ref{normal}), in red the region corresponding to (\ref{anom}).}
\label{fig_stab}
\end{figure}

In this section we
will focus on the simplest case where the canonical ${ D}$ and ${  M}$ matrices are {\it time independent}. As we already discussed, this
will be the case when shift symmetry is imposed for all the fields
and the metric is time-independent. 
The equations of motion are the
following%
\be
\ddot \varphi {  -} 2\;{  D} \, \dot \varphi+{ 
  M}\,\varphi=0 \, .
\label{eqm}
\ee
Solutions of (\ref{eqm}) are of the form $\varphi= e^{-i\,\omega\,t}\, v $ where $v$ is
suitable vector and $\omega$ satisfies the following algebraic
equation given in terms of the linear operator $L(\omega)$
 \be
 \det L \left(\omega \right) \equiv \det \left (-\omega^2\,{\bf
     I}+2\,i\,\omega \, {  D}+{  M} \right)=0
 \quad \Rightarrow \quad \omega ^4 - \omega ^2 \left(4 \, d^2
   +m_1^2+m_2^2\right)+m_1^2\,m_2^2=0 \, ;
 \label{eqomega}
 \ee
thus
\be
\omega_{1,2}^2=\frac{1}{2} \left(4\, 
   d^2+m_1^2+m_2^2\pm \sqrt{\left(m_1^2 +m_2^2 +4 \, d^2\right)^2 - 4
     \, m_1^2 \, m_2^2} \right) \, .
\label{omdef}
\ee
Some simple general properties of $\omega_i$ can obtained by taking
the transpose and then the complex conjugate of  $L(\omega)$ and
taking into account that ${ D}^t = -{  D}$.
In particular we  get $L(-\omega)=L(\omega)$ and 
$L(-\omega^*)=L(\omega)^*$.
Thus, if $\omega$ is a solution, also $\omega^*$, $-\omega$  and
$-\omega^*$ are solutions. The system   is stable only when $\omega$
is purely real. 
The region of stability can be described by  using  the  $m_{1,2},
\;d$ or equivalently $\omega_{1,2}, \;d$ as independent parameters and it is given by
\bea
(a) && m_{1,2}^2\geq0,\;\;d^2>0\hspace{ 1cm} \Longleftrightarrow
\hspace{ 1cm} 0\leq d^2\leq \frac{\left(\omega_1-\omega_2\right)^2}{4}
\, ; \label{normal}\\
 (b) && m_{1,2}^2\leq0,\;\;d^2\geq \frac{\left(\sqrt{-m_1^2}+\sqrt{-m_2^2}\;\right)^2}{4} ,\quad 
 \Longleftrightarrow \quad  d^2>
 \frac{\left(\omega_1+\omega_2\right)^2}{4} \;\, .
\label{anom}
\ea
One can also rescale $\omega_i$ by $d$ by defining $\hat\omega_i\equiv
\omega_i/d$, then the two stability regions corresponding to
$\hat\omega_1\geq\hat\omega_2\geq0$ can be rewritten as
 \bea\label{rega}
&&(a)\quad \hat \omega_1-\hat\omega_2\geq 2
\\\label{regb}
&& (b)\quad 
 \hat \omega_1+\hat\omega_2\leq 2 \, .
\ea
Our notion of stability corresponds to what   it is called
marginal stability in~\cite{Lancaster}. The region of stability in
parameter space is plotted in Figure \ref{fig_stab}.
The intermediate range $\frac{\left(\omega_1-\omega_2\right)^2}{4}<d^2<
\frac{\left(\omega_1+\omega_2\right)^2}{4}$ is forbidden by stability.
We note~\footnote{We thank the anonymous referee for pointing out
  such a limit.} that stability can be also achieved when the rather peculiar
limit  ${\cal K} \to 0$ in (\ref{quadlc}) is taken; namely the
standard kinetic term which gives rise in the canonical form to
$\ddot{\phi}$ is sub-leading when compared to the gyroscopic one. In
such a limit, from (\ref{eqm}) one find that $\omega^2=m_1^2 \,
m_2^2/(4 d^2)>0$. Of course by taking the limit  ${\cal K} \to 0$ the
number of degrees of freedom is changed and the equations of motion
are not anymore second order differential equations; see
\cite{Adshead:2016iix} for an application to inflation.

It is interesting to interpret the stability conditions in
Hamiltonian terms. The conjugate momenta are
\be
\pi = \dot{\varphi} - { D} \, \varphi \, ,\quad
\pi= 
\begin{pmatrix}
\pi_1 \\
\pi_2
\end{pmatrix} \, .
\label{cjm}
\ee
Working in Fourier space, the Hamiltonian $H$ can be written as 
%
\bea
  H = \int
d^3k \, H_k  
\!\!&=&\!\! \int
d^3k \, \left[\ha \left[  \pi^\dagger_{\pmb{k}} \, \pi_{\pmb{k}} +
\varphi^\dagger_{\pmb{k}} \, ( {  M}-{   D}^2)\, \varphi_{\pmb{k}} \right]+
\pi^\dagger_{\pmb{k}}   \,  {  D } \,\varphi_{\pmb{k}}\right] =
\\&&\nonumber
\int
d^3k \, \ha \left[  (\pi_{\pmb{k}}+{   D} \,\varphi_{\pmb{k}})^\dagger (\pi_{\pmb{k}}+  {  D}\,\varphi_{\pmb{k}}) + 
    \varphi^\dagger_{\pmb{k}}  \,{  M}\, \varphi_{\pmb{k}} \right]\,  .
    \label{ham}
\ea
The system is stable in two disconnected regions in parameter
space. 
\\
The first one, that we call {\it normal}, is given by (\ref{normal})
and the Hamiltonian ${ H}_k$ is positive defined as one can infer
from (\ref{ham}). 
More surprising is the existence of a second
 region (\ref{anom}) (that we call {\it anomalous region}) where  ${ H}_k$ is not positive
 defined, but the system is still stable. 
 Thus the positivity of
 ${H}_k$ is only a sufficient but not necessary condition for
 stability. Notice that the stability in the anomalous region is possible only
 if ${  D} \neq 0$.
  We have focused on the case of
 two degrees of freedom, but more in general  one can show that
 stability in the anomalous region can be achieved if and only if the number of degrees of
 freedom is even and by taking  ${  D}  $  non-singular and
 sufficiently large \cite{article}; indeed the
 generalization of (\ref{omdef}) leads to
 \be
4\,{  D}^t \, {  D} -\left(\sqrt{-{ M} }+\sqrt{{ 
      D}^t\, {  M} \, {  D}^{-1}}\right)^2>0 \, .
\ee
Another surprising property is that even if ${  D}$ and ${  M}$
are time-independent,  taking  $\varphi(t)$ as a solution of
(\ref{eqm}), then
$\tilde \varphi(t) ={\bf T }\, \varphi(-t)$ is not a solution for any choice
of a constant 2$\times$2 matrix ${\bf T}$ unless ${  D} =0$. In other words in a
gyroscopic system time reversal symmetry is broken even if the energy is
conserved. The reason behind the lack of time reversal symmetry is the
form of (\ref{cjm}) that generates an Hamiltonian that is not an even function of the conjugate momenta $\pi$. For a recent critical discussion of time reversal symmetry
see~\cite{Roberts:2016ofs}.
\\
It is important to say some words about the effect of dissipation on a gyroscopic system where a lot of literature is present (see \cite{Krechetnikov:2007zz}).
According to the Thomson-Tait-Chetayev  theorem, 
for a gyroscopic system defined in the region \ref{anom} with a ``large" $D$, stability is destroyed by the introduction of an arbitrarily small dissipative force.
\section{Symplectic Classical Dynamics}
As discussed in the previous section the impossibility to get rid of
${  D}$ at the Lagrangian level makes the Hamiltonian formalism the
ideal tool both to study the classical dynamics and to quantize the
system. 
Following the approach described in~\cite{Grain:2019vnq,Colas:2021llj}, it is convenient to introduce a compact notation to denote a generic
point in the phase space  described by a 4-dimensional vector $z(t,\vec x)$ 
or equivalently its Fourier transform $z_{\pmb{k}}(t)$ defined by
\be
z_{\pmb{k}}= \begin{pmatrix}
\varphi_{\pmb{k}} \\
\pi_{\pmb{ k}}
\end{pmatrix} \, , \qquad
 \varphi= 
\begin{pmatrix}
\varphi_1 \\
\varphi_2
\end{pmatrix} 
,\quad
\pi= 
\begin{pmatrix}
\pi_1 \\
\pi_2
\end{pmatrix} \, .
\label{not}
\ee
The Hamiltonian can be written as (for quadratic systems see also \cite{PhysRev.139.A450}, \cite{PhysRevA.99.022130}, \cite{PhysRevA.103.023707})
\be
H = 
\int d^3k\;\frac{1}{2}\;z^t_{-\pmb{k}} \,  {\cal H}_k \, z_{\pmb{ k}}=
\int d^3k\;\frac{1}{2}\;z^\dagger_{\pmb{ k}}\,  {\cal H}_k \, z_{\pmb{ k}} 
\label{hamint}
\ee
The Hamiltonian density matrix in Fourier space can be read off  from (\ref{quadlc}) and (\ref{ham}) in the canonical form as
\be
{\cal H}_k =\,
\begin{pmatrix}
{  M} -{ D}^2 & { D}^t\\ 
{ D} & \pmb{I} 
\end{pmatrix}
=  
\begin{pmatrix}
 m_1^2+d^2 & 0& 0&- d\\  
0&  m_2^2+d^2& d& 0\\ 
0& d& 1& 0\\   
 -d & 0& 0&1\\  
\end{pmatrix}
\label{Hmat}
\ee
with $k =|\pmb{ k}|$. 
The Poisson brackets among the basic variables can be
written as
\be
\{  z_m(t,\pmb{x}) ,\,  z_n(t,\pmb{ y}) \}=\Omega_{mn} \, \delta^{(3)} (\pmb{ x}-\pmb{ y}),\quad \Rightarrow\quad
\{z_{\pmb{ k},m}(t ) ,\, z_{\pmb{ q},n}^\ast(t ) \}=\Omega_{mn} \,
\delta^{(3)} (\pmb{k}-\pmb{ q}) \, ,
\label{pb}
\ee
where $\Omega$ is the following 4$\times$4 antisymmetric matrix that encodes the symplectic structure
\be
\Omega=
\begin{pmatrix}

0 & \pmb{I}\\
- \pmb{I} & 0
\end{pmatrix} \, .
\ee
The Hamilton equations can be written in terms of the Poisson
brackets as a set of linear first order differential equations
\be
\dot z_{\pmb{k}}(t )=\{z_{\pmb k}(t ),\;H\}=\Omega\; {\cal H}_k \,
z_{\pmb k}(t)
\label{Heq}
\ee
that are equivalent to (\ref{eqm}). From now on for notation simplicity
we will omit the suffix $\pmb{k}$ in $z_{\pmb k}$.
We exploit the freedom in the choice of canonical variables to find a
symplectic transformation\footnote{According to the  Williamson
theorem, given a positive definite and symmetric
Hamiltonian  $H$, it always exists a  symplectic
transformation that diagonalises $H$; see appendix~\ref{hamdia}
and~\cite{Nicacio:2021zmc} for a recent discussion.} that diagonalises
the Hamiltonian ${\cal H}_k $
\be
S^t\,  {\cal H}_k \, S= \Lambda_{\cal H}
\label{sh}
\ee
where $\Lambda_{\cal H}$ is a diagonal matrix  and the symplectic
matrix $S$, namely
\be
S^t \, \Omega \, S =  \Omega \, .
\ee
The symplectic decomposition (\ref{sh}) is different
from a similarity transformation used in the standard diagonalization procedure.
Once $S$ is found, time evolution is rather simple in the new basis
$\tilde z$ defined by $z=S \,  \tilde z$ where the system can be interpreted as a collection of decoupled
harmonic oscillators and then quantization becomes standard.
To find  $S$ we consider the following ansatz 
\be
S=\begin{pmatrix}
 \pmb{I} &B \\C & J 
\end{pmatrix} \, ,\qquad J=\begin{pmatrix}
j_{11}&0\\0 &j_{22}
\end{pmatrix} \, ,\qquad C=\begin{pmatrix}
0&c\\c&  0
\end{pmatrix}\, ,\qquad B=\begin{pmatrix}
0&b\\b& 0
\end{pmatrix} \, .
\ee
Imposing that $S$ is symplectic produces a 2-parameter family of
symplectic matrices; the parameters $b$ and $c$ can be fixed by
imposing that   $\Lambda_{\cal H}$ is diagonal and one gets
\be
\Lambda_{\cal H} = 
\begin{pmatrix}
  \frac{ 2\;\omega _1^2\;\left(\omega _1^2-\omega _2^2\right)}{T_{11}} & 0 & 0 & 0 \\
 0 & \frac{ 2\;\omega _2^2\;\left(\omega _1^2-\omega _2^2\right)}{T_{22}} & 0 & 0 \\
 0 & 0 &\frac{T_{11}}{2 \,(\omega _1^2-  \omega _2^2)} & 0 \\
 0 & 0 & 0 &  \frac{T_{22}}{2 \,(\omega _1^2- \omega _2^2)}\\
\end{pmatrix} \, ;
\ee
where\footnote{Note that 
 $m_1^2-m_2^2=\sqrt{(4\,d^2-\o_1^2-\o_2^2)^2-4\,\o_1^2\,\o_2^2}$ and 
 $\o_1^2-\o_2^2=\sqrt{(4\,d^2-m_1^2-m_2^2)^2-4\,m_1^2\,m_2^2}$,
that allow to express all the quantities as functions of the independent
parameters $d,\,\o_{1,2}$ or $d,\,m_{1,2}$.}
\be
\begin{split}
  & T_{11}=4 \, d^2 +m_1^2-m_2^2+ \omega_1^2-\omega_2^2 \, , \quad
  T_{22}=-4 \, d^2 +m_1^2-m_2^2+ \omega_1^2-\omega_2^2 \, ,\\
  & T_{33}=4 \, d^2 -m_1^2+m_2^2+ \omega_1^2-\omega_2^2 \, , \quad
   T_{44}=-4 \, d^2 -m_1^2+m_2^2+ \omega_1^2-\omega_2^2  \, ;
\end{split}
\ee
and
\be
j_{11} = j_{22}= 1+ b \; c \, , \qquad  c=\frac{m_1^2-m_2^2- \omega_1^2+ \omega _2^2}{4 \;d} \, , \qquad b=
\frac{2\; d}{ \omega _1^2- \omega _2^2} \, .
\ee
Notice also that
\be
T_{11}\;T_{33}=16\;d^2\;\omega_1^2\to\Lambda_{{\cal H}{}_{11}} \, \Lambda_{{\cal H}{}_{33}} = \frac{\omega_1^2}{4} \, ,
\qquad T_{22}\;T_{44}=16\;d^2\;\omega_2^2\to\Lambda_{{\cal H}{}_{22}} \, \Lambda_{{\cal H}{}_{44}} =
\frac{\omega_2^2}{4} \,.
\label{rel}
\ee
The system is classically stable when $\omega_{1,2}\in{\bf R}$ and $\omega_1^2\geq\omega_2^2\geq 0$;
however as discussed in the previous section stability does not
implies that the Hamiltonian is positive definite.  Indeed 
\be
\begin{split}
\label{TT}
& T_{11}\geq 0 \;{\rm and}\;T_{33}\geq 0 \;\;{\rm for}\;\; 0\leq
d^2\leq\frac{(\omega_1-\omega_2)^2}{4}
\; \text{and} \;d^2\geq\frac{(\omega_1+\omega_2)^2}{4} \, ; \\
& T_{22}\geq 0\;{\rm and}\;T_{44}\geq 0 \;\;{\rm for}\;\; 0\leq
d^2\leq\frac{(\omega_1-\omega_2)^2}{4}  \, .
\end{split}
\ee
For $0\leq d^2\leq\frac{(\omega_1-\omega_2)^2}{4}$  the diagonal
Hamiltonian is positive defined and corresponds to the normal region
(\ref{normal}) of stability.
\\
 In the anomalous region (\ref{anom}),
$d^2\geq\frac{(\omega_1+\omega_2)^2}{4}$ and the system is still stable,
but now $T_{11,}, \, T_{33} >0$ while  $T_{22,}, \, T_{44} <0$ and the
Hamiltonian  can be written as the sum of  one standard harmonic oscillator
plus a  second ghost-like  harmonic oscillator.
\\
This can be shown explicitly by exploiting the fact that we  can
still perform a further canonical transformation to reduce the
oscillators Hamiltonian to the standard form
\be
\tilde z= N_{\pm} \,  z_c\,, \qquad N_{\pm}=\frac{1}{\sqrt{2}} \begin{pmatrix}
n_{\pm}&   0\\
  0  & n^{-1}_\pm
\end{pmatrix} \, ;
\ee
exploiting (\ref{rel}), the 2$\times$2 submatrix is taken as
 \be
 n_\pm =  
\begin{pmatrix} \sqrt{ \frac{T_{11} }{ 2 \,\omega_1\; (\omega_1^2-\omega_2^2) }}
& 0 \\
0 & \sqrt{\frac{ \pm T_{22}}{2 \,\omega_2 \;(\omega_1^2-\omega_2^2)}}\\
\end{pmatrix} \, ;
\label{nexpr}
\ee
the notation  $\pm$ refers to the case where $T_{22}$  is positive
or negative according to (\ref{TT}). The integrand that
defines the Hamiltonian in (\ref{hamint}) reads
\be
\label{Hc}
\Lambda_c^{(\pm)}=N^t_\pm \,  {\Lambda}_{\cal H} \,  N_\pm= \begin{pmatrix}
 \omega_1& 0&0&0 \\
0&\pm\,\omega_2 &0&0 \\ 
0&0&\omega_1& 0 \\
0&0&0&\pm\,\omega_2 \\
\end{pmatrix}\,  , \qquad H_k^{(\pm)}= \ha \, z_c^\dagger\;\Lambda_c^{(\pm)}\;z_c \, .
\ee
The explicit expressions for the Hamiltonian is then given by
\be
H_k^{(+)}=\ha \sum_{i=1,2}\omega_i\;\left(\pi^2_{c{}_i}+\varphi^2_{c{}_i}\right),\qquad
H_k^{(-)}=
\frac{\omega_1}{2}\,\left(\pi^2_{c{}_1}+\varphi^2_{c{}_1}\right)-\frac{\omega_2}{2}
\, \left(\pi^2_{c{}_2}+\varphi^2_{c{}_2}\right)
\, .
\ee
The quantization of the system in the anomalous region that
corresponds to  $H_c^{(-)}$ has a 
ghost character. The complete canonical transformation that relates the
original variables $z$ and $z_c$ is given by
\be
z=S \,  \tilde{z} = S \,  N_\pm \, z_c \, , \qquad \Lambda_c^{(\pm)}=
N_\pm^t \,  S^t\,  {\cal H} \, S \, N_\pm \, .
\label{zt}
\ee
The time evolution of  $z_c$ is very simple 
\be
\begin{split}
&z_c(t)=e^{\Omega\, \Lambda_c^{(\pm)}\,t}\,  z_c(0) \equiv G_c^{(\pm)}(t) \,
z_c(0) \, , \\
&G_c^{(\pm)}(t)=\begin{pmatrix}
\cos \left(t \, \omega_1\right) & 0 & \sin \left(  t \, \omega_1\right )& 0 \\
 0 & \cos \left(t \, \omega_2 \right) & 0 & \pm \sin
 \left(t \, \omega _2 \right) \\
 - \sin \left(t \, \omega_1 \right) & 0 & \cos\left(t
     \, \omega _1\right) &   0 \\
 0 & \mp \sin \left(t\, \omega_2\right) & 0 & \cos
 \left(t \, \omega_2\right)
  \end{pmatrix} \, .
\end{split}
\ee
The matrix $G_c^{(\pm)}$ is also symplectic.
From $G_c^{(\pm)}$ the evolution of the original variables $z(t) $ can be also
found by using (\ref{zt}):
\be
z(t) = { G}(t) \, z(0)\, , \qquad G(t) = S \, N_\pm \, G_c^{(\pm)}(t) \,
N^{-1}_\pm \, S^{-1} \, .
\label{zevol}
\ee

\section{Quantization}
\begin{figure}[htbp]
\centering
\includegraphics[width=0.4 \columnwidth]{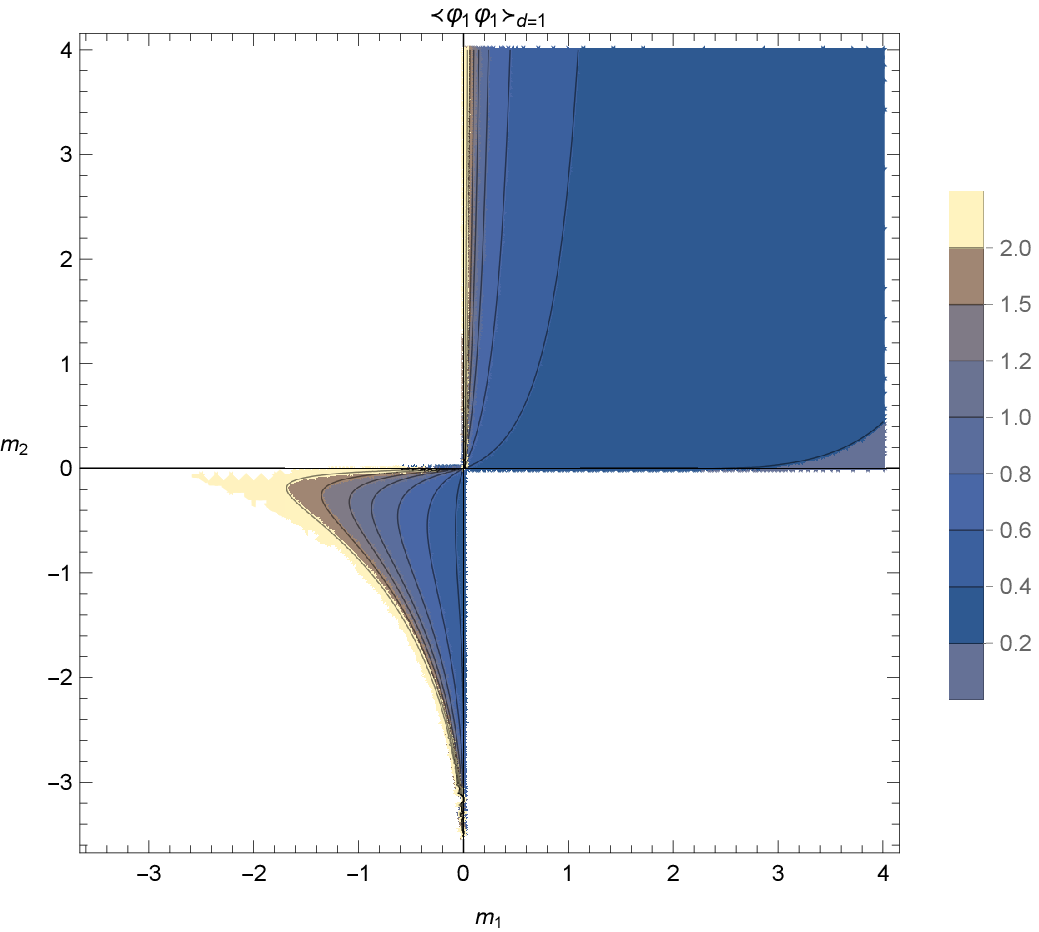}
\hspace{1.cm}
\includegraphics[width=0.4 \columnwidth]{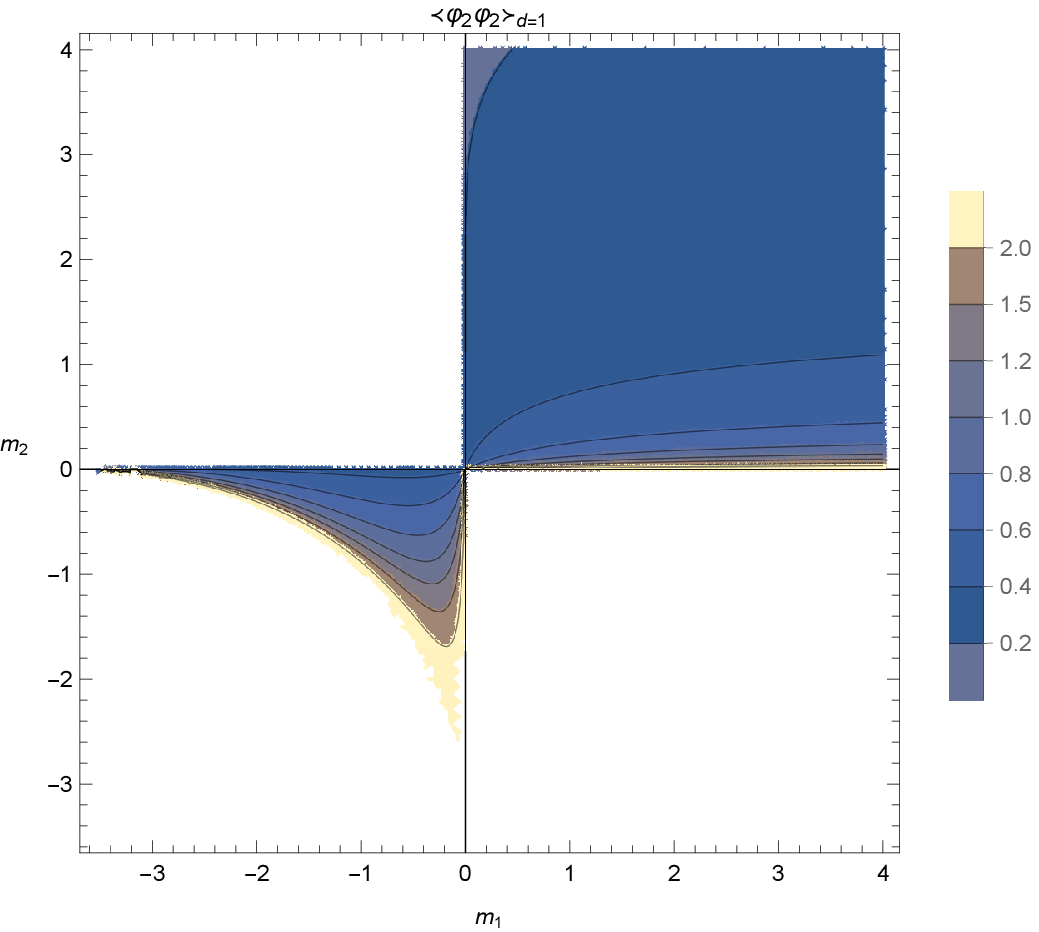}
\caption{Contour plot of ${\cal Z}_{11}$ and ${\cal Z}_{22}$ for $d=1$.}
\label{PS_orig}
\end{figure}

\begin{figure}[htbp]
\centering
\includegraphics[width=0.6\columnwidth]{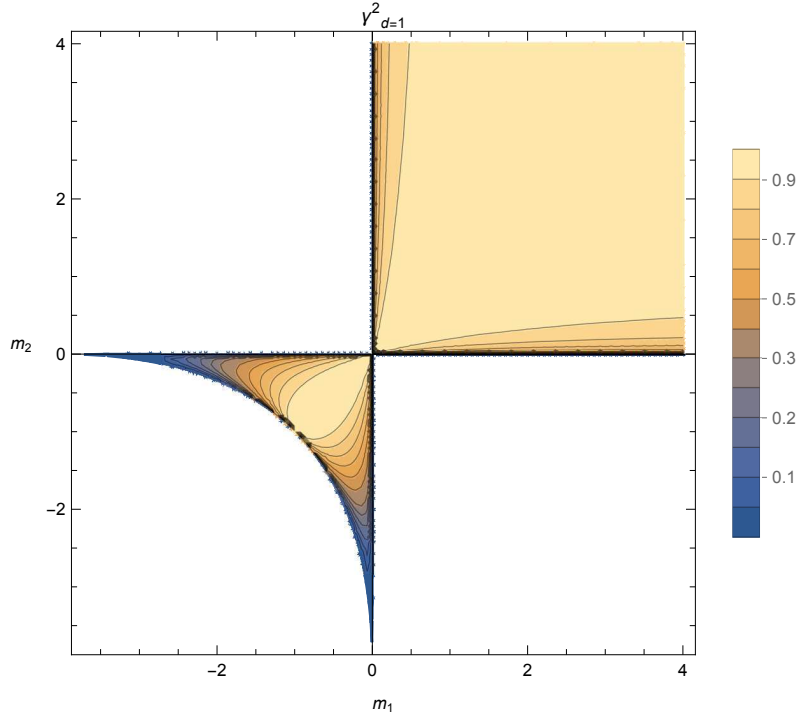}
\caption{Contour plot  of the square for the purity $\gamma^2$  once we fixed $d=1$.}
\label{figure4}
\end{figure}

One of the problem in the quantization of  classical field theory is
that, contrary to the case of system with a finite number of degrees of
freedom, the procedure is not unique~\cite{Haag:1992hx}; indeed, given two representations of the canonical commutation
relations, in general it is not
guaranteed that they are unitary equivalent. The most widely used
quantization scheme is based on the Fock space construction according with
when a suitable set of  creation and annihilation operators are defined,
physical states are built by acting with them on the vacuum state.  While in flat
spacetime Poincare' symmetry allows a natural selection of the vacuum
state, in general this is not the case and different set of creation
and annihilation operators can be constructed related by a Bogolyubov
transformation. A well known example is the study of quantum field in
a non-trivial gravitational background~\cite{Birrell:1982ix,Wald:1995yp,Mukhanov:2007zz,Parker:2009uva}. Typically
the first step is to write the non-interacting Hamiltonian of the
system  as a set of decoupled  harmonic oscillators; given a quadratic
Hamiltonian, one can introduce creation and annihilation operators
starting from a set of canonical variables $z_{\pmb{k}}$  and the
classical Poisson brackets by promoting them to quantum operators ($z_{\pmb{k}}\to \hat z_{\pmb{k}}$) which
satisfy the equal time canonical commutation relations (CCR)~\cite{Dirac-pqm} (see \cite{Grain:2019vnq,Colas:2021llj} for notations and method)
\be
[ \hat z(t,\pmb{x})_m,\,  \hat z(t,\pmb{ y})_n]
= i \, \Omega_{mn} \, \delta^{(3)} (\pmb{ x}-\pmb{ y}) \, .
\ee
The easiest way to construct a Fock representation of the CCR is to
start from the diagonal form of the Hamiltonian (\ref{Hc})  in terms of canonical
variables $z_c$. In the Heisenberg picture we define
\be\label{defb}
b_{{\pmb{k}}{}_j}(t) = \frac{1}{\sqrt{2}} \left(
    \hat\varphi_{\pmb{k}}{}_{c_ j} +  i\;
   \hat \pi_{\pmb{k}}{}_{c _ j}  \, \right) \, , \qquad
    b_{{\pmb{k}}{}_j}^\dagger(t) = \frac{1}{\sqrt{2}} \left(
    \hat\varphi_{\pmb{k}}{}_{c_j} - i \,  
    \hat\pi_{\pmb{k}}{}_{c _ j}  \, \right)  \quad j=1,2 \, ,
\ee
which leads to
\be
\left [b_{{\pmb{k}}{}_m}(t), \, b^\dagger_{{\pmb{q}}{}_n}(t) \right]=\delta_{m n}
  \delta^{(3)}(\pmb{k}-\pmb{q}) \, .
\ee
In an equivalent and more compact matrix notation
\be
B_{\pmb{k}} (t)= U \, \hat z_{{\pmb{k}}{}_c}(t) \, ;
\label{bdef}
\ee
where
{ 
\be
\begin{split}
  &B_{\pmb{k}}(t) = \left(b_{{\pmb{k}}{}_1}(t), \, b_{{\pmb{k}}{}_2}(t),
  b_{{\pmb{-k}}{}_1}^\dagger(t), \, b_{{\pmb{-k}}{}_2}^\dagger(t) \,
\right)^t \, , 
\qquad  U = \frac{1}{\sqrt{2}}\begin{pmatrix}  \pmb{I}&  i \, \pmb{I} \\
 \pmb{I}  & -i\,  \pmb{I} \end{pmatrix} \, ; 
\end{split} 
\ee
}
with the corresponding inverse relation
\be
\hat z_{{\pmb{k}}{}_c}(t) = U^\dagger \; B_{\pmb{k}}(t)  \, .
\ee
It is easy to show that
\be
b_{{\pmb{k}}{}_j}(t)= e^{ -i \, \omega_j \, t} \, b_{{\pmb{k}}{}_j} \, ,
\qquad \qquad j=1,2 \, .
\label{btevolv}
\ee
The vacuum state relative to the $b$ operator is defined as
\be
b_{{\pmb{k}}{}_j}(t_0)\, |0_b \rangle =0  \, ,\qquad \langle 0_b |0_b
\rangle =1\;.
\label{bdeft0})
\ee
Actually, given the time evolution (\ref{btevolv}), setting $t_0=0$,
(\ref{bdeft0}) holds for any $t$.
The Hamiltonian in terms of these creation and annihilation operators has the
standard form for two independent harmonic oscillators
\be
H_k^{(+)}=\sum_{i=1}^2 \frac{\omega_i}{2} \left(b_{{\pmb{k}}{}_i} \,
  b_{{\pmb{k}}{}_i}^\dagger +  b_{{\pmb{k}}{}_i}^\dagger \,
  b_{{\pmb{k}}{}_i} \right)
\label{hnorm}
\ee
in the normal region and
\be
H_k^{(-)}=\frac{\omega_1}{2} \left(
b_{{\pmb{k}}{}_1} \,
  b_{{\pmb{k}}{}_1}^\dagger +  b_{{\pmb{k}}{}_1}^\dagger \,
  b_{{\pmb{k}}{}_1} \right)- \frac{\omega_2}{2} \left(b_{{\pmb{k}}{}_2} \,
  b_{{\pmb{k}}{}_2}^\dagger +  b_{{\pmb{k}}{}_2}^\dagger \,
  b_{{\pmb{k}}{}_2} \right) \,
\ee
in the anomalous region of stability.
The correlation function for canonical fields can be  easily obtained \cite{Grain:2019vnq,Colas:2021llj} as
\be
\begin{split}
 \langle 0_b| \hat z_{\pmb{k}}{}_{c{}_m}(t) \, \hat z_{\pmb{q}}{}_{c{}_n}^\dagger(t)   |0_b
\rangle&=  U^\dagger  _{mr} \, \langle 0_b|  B_{{\pmb{k}}{}_r}(t) \,  B_{{\pmb{q}}{}_s}(t)  |0_b
\rangle\,  U_{sn}  \\
&= \delta^{(3)}(\pmb{k} - \pmb{q}) \;
U^\dagger_{mr}  \;  
\left(  \delta_{r\,1} \,  \delta_{s\,3} +\delta_{r\,2} \,  \delta_{s\,4} \right)\;  
  U_{sn}
\equiv \delta^{(3)}(\pmb{k} - \pmb{q}) \; \Sigma_{mn}  \\[.2cm]
& \Sigma= \ha \left( \pmb{I} +i \, \Omega\right)= \begin{pmatrix}
\frac{1}{2 } & 0 & \frac{i}{2} & 0 \\
 0 & \frac{1}{2 } & 0 & \frac{i}{2} \\
 -\frac{i}{2} & 0 & \frac{1}{2} & 0 \\
 0 & -\frac{i}{2} & 0 & \frac{1}{2}
\end{pmatrix}
\end{split}
\ee
The same correlation function for the original fields can be also
computed by using (\ref{zt}), namely  $\hat z_{\pmb{k}}(t)= S \, N_\pm \, \hat z_{\pmb{k}}{}_c(t)$
\be
\begin{split}
&\langle 0_b| \hat z_{{\pmb{k}}{}_m}(t) \, \hat z_{{\pmb{q}}{}_n}^\dagger(t)   |0_b \rangle=
\delta^{(3)}(\pmb{k} - \pmb{q}) \, Z_{mn} \, ; \\[.2cm]
& Z= S \, N_\pm \, \Sigma \, N_\pm^t \, S^t \,  \, \equiv {\cal Z}+\frac{i}{2}\;\Omega;
\end{split}
\label{corr}
\ee
with $Z$ hermitian matrix with entries: 
 \be
 {\cal Z} = \frac{1}{(\omega_1^2-\omega_2^2)} \begin{pmatrix} \frac{4 \,d^2
     \, \omega_2}{ \pm \,T_{22}  }+\frac{T_{11}}{4\, \omega_1} & 0 & 0
   & {\cal Z}_{14} \\
   0 & \frac{4 \,d^2 \,\omega_1}{T_{11}}+\frac{\pm \,T_{22}  }{4 \,\omega_2} &
   {\cal Z}_{23}  & 0\\
   0 &  {\cal Z}_{23} &  {\cal Z}_{33} & 0\\
   {\cal Z}_{14} & 0 & 0 & {\cal Z}_{44} 
\end{pmatrix} \, ;
\ee
where
\be
\begin{split}
& {\cal Z}_{14} =  \frac{T_{11}
  \left(m_1^2-m_2^2-\omega_1^2+\omega_2^2\right)}{16 \, d \, 
  \omega_1}+\frac{d \, \omega_2
  \left(m_1^2-m_2^2+\omega_1^2-\omega_2^2\right)}{\pm\, T_{22}} \, ;\\
&{\cal Z}_{23}= \frac{d \, \omega_1
     \left(m_1^2-m_2^2+\omega_1^2-\omega_2^2\right)}{T_{11}}+\frac{\pm \,T_{22}
   \left(m_1^2-m_2^2-\omega_1^2+\omega_2^2\right)}{16 \, d
 \,   \omega_2} \, ;\\
 &{\cal Z}_{33}= \frac{1}{64} \left[\frac{\pm \,T_{22}
   \left(m_1^2-m_2^2-\omega_1^2+\omega_2^2\right){}^2}{d^2\,
   \omega_2}+\frac{16\, \omega_1\,
   \left(m_1^2-m_2^2+\omega_1^2-\omega_2^2\right){}^2}{T_{11}}\right]
\, ;
\\
& {\cal Z}_{44}= \frac{1}{64} \left(\frac{T_{11}
    \left(m_1^2-m_2^2-\omega_1^2+\omega_2^2\right){}^2}{d^2 \, \omega_1}+\frac{16
   \, \omega_2 \left(m_1^2-m_2^2+\omega_1^2-\omega_2^2\right){}^2}{\pm\,
   T_{22} }\right) \, ;
 \end{split}
 \ee
 While both $\Sigma$ and $Z$ are symplectic matrices with the same
 symplectic eigenvalues, the canonical transformation (\ref{nexpr})
 is singular in the limit $\omega_2 \to \omega_1$ and it is interesting
to see  what happens to the correlation. From (\ref{omdef}), $\omega_2
= \omega_1$ is possible when
\be
\begin{cases} d = d_{\text{cr}} =
\frac{\left(\sqrt{-m_1^2}+\sqrt{-m_2^2}\;\right)}{2} & \text{anomalous
  region } m_{1,2}^2 <0\\[.2cm] 
d=0, \; m_2^2=m_1^2 >0 & \text{normal region }  m_{1,2}^2 >0
\end{cases} \, .
\ee
In the anomalous region when $d=d_{\text{cr}}$, we have that
  $\omega_1^2=\omega_2^2=\bar \omega^2 \equiv \left( m_1^2 \, m_2^2\right)^{1/2}$, then setting
   $ d=d_{\text{cr}}+ \frac{\epsilon^2}{2}$  it gives
   $\omega_1^2=\bar \omega^2+ 2 \, \bar \omega \, d_{\text{cr}} \, \epsilon$,
   $\omega_2^2=\bar \omega^2- 2 \, \bar \omega \, d_{\text{cr}} \, \epsilon$,
   where $\epsilon$ is a dimensionless small parameter measuring the
   distance of $d$ from $d_{\text{cr}}$.
   The non-trivial part ${\cal Z}$ of the  correlation function
   (\ref{corr}) has the following behavior  at the leading order in
   $\epsilon$
   \be
   {\cal Z} = \begin{pmatrix}
     \frac{1}{2 \, \epsilon \, \sqrt{-m_1^2}} & 0 & 0
     &\frac{\sqrt{-m_2^2}-\sqrt{-m_1^2}}{4 \, \epsilon \,
       \sqrt{-m_1^2}}\\
     0 &  \frac{1}{2 \, \epsilon \, \sqrt{-m_2^2}} &
 \frac{\sqrt{-m_1^2}+\sqrt{-m_2^2}}{4 \, \epsilon \, \sqrt{-m_2^2}} &
 0\\
 0 &  \frac{\sqrt{-m_1^2}+\sqrt{-m_2^2}}{4 \, \epsilon \,
   \sqrt{-m_2^2}} &\frac{\left(m_1^2-m_2^2\right){}^2}{8   \, \epsilon \,\sqrt{-m_2^2}
   \left(\sqrt{-m_1^2}+\sqrt{-m_2^2}\right){}^2} & 0\\
 \frac{\sqrt{-m_2^2}-\sqrt{-m_1^2}}{4 \, \epsilon \,
       \sqrt{-m_1^2}} & 0 & 0 & \frac{\left(m_1^2-m_2^2\right){}^2}{8
       \, \epsilon \, \sqrt{-m_1^2}
   \left(\sqrt{-m_1^2}+\sqrt{-m_2^2}\right){}^2}\\
   \end{pmatrix} \, .
   \ee
 Thus $ {\cal Z} $ shows a resonant singular behavior  when $\omega_1
 \approx \omega_2$ in the anomalous region of stability, which is
peculiar also from a classical point view: even a very small coupling can trigger a
runaway behavior  of classical solutions that far from
$\omega_1 \approx \omega_2$ are well behaved at least when interactions
are not too big~\cite{Gross:2020tph}. 
On the other hand   when $d \to 0$ in the normal region of stability
no resonant behavior  is present; indeed
we get
 \be
 {\cal Z} = \begin{pmatrix}
   \frac{1}{2 \, m_1} & 0 & 0 &0\\
   0 &  \frac{1}{2 \, m_1} & 0 & 0\\
   0 & 0 & \frac{ m_1}{2 } & 0 \\
   0 & 0 & 0 & \frac{ m_1}{2 }
    \end{pmatrix} \, .
   \ee
The entries $Z_{11}$ and $Z_{22}$ are particularly important, since they represent the autocorrelations ( power spectra) of the original fields $\vf_1$ and $\vf_2$, whose behavior  in both the stability regions is shown in figure \ref{PS_orig}.
 \\
From the fact that $S$ and $N_\pm$ are symplectic they have unit
determinant, we get
\be
\text{Det} \left( Z \right)=\text{Det} \left( \Sigma\right)
\Rightarrow \text{Det} \left({\cal Z} \right)=\frac{1}{16} \, .
\ee
As shown in \cite{Colas:2021llj} the part ${\cal Z}$ of the
correlation matrix $Z$ is relevant to study decoherence once one of
two modes is traced out.
In particular, the so called {\it purity} $\gamma$ is a measure of the
entanglement between the two dof.
%
\be
\gamma^2\equiv \left[4\;({\cal Z}_{11}\;{\cal Z}_{33}-{\cal
    Z}^2_{13}) \right]^{-1} \, .
\ee
In particular for a pure state we have $\gamma=1$ while for mixed states we have $0\leq \gamma\leq 1$. The limit  $\gamma\to 0$ corresponds to the maximally decoherence case \cite{Colas:2021llj}.
For our gyroscopic system in particular we get in the two regions of stability the following results
\be
 \gamma^2_{\pm}=
 \frac{\hat{\omega }_1
  \left(\hat{\omega }_1\pm\hat{\omega }_2\right){}^2 \hat{\omega }_2}{4
   \left[\left(\hat{\omega }_1\pm\hat{\omega
   }_2\right){}^2-4\right] \left(\hat{\omega }_1
   \hat{\omega }_2\pm1\right)} \, ,
\ee
that we show in figure \ref{figure4}. We give also the following limits on the boundaries of the parameter space
\bea
&& \lim_{d\to 0}\gamma^2_{+}=1,\qquad \lim_{d\to \infty}\gamma^2_{+}=\frac{4\,\sqrt{m_1\,m_2}}{(\sqrt{m_1}+\sqrt{m_2})^2}
 \\
 && \lim_{d\to d_c}\gamma^2_{-}=0,\qquad \lim_{d\to \infty}\gamma^2_{-}=\frac{4\,\sqrt{m_1\,m_2}}{(\sqrt{-m_1}+\sqrt{-m_2})^2}\\
 &&
  \lim_{m_1\to m_2}\gamma^2_{\pm}=1
 \ea

As discussed at the beginning of the section the Fock representation is
by non means unique. Indeed given a set a canonical variable one can
define an alternative set of creation/annihilation operators by
applying the above construction for $\hat z_c$ to $\hat z$. Similarly to
(\ref{defb}) we can define a new set of creation/annihilation operators
$a$ and $a^\dagger$ such that
\be
a_{{\pmb{k}}{}_j}(t) = \frac{1}{\sqrt{2}\;y_j} \left(y_j^2 \, 
    \hat\varphi_{\pmb{k}}{}_{ j} +  i\,
   \hat \pi_{\pmb{k}}{}_{  j} \, \right) \, , \qquad
    a_{{\pmb{k}}{}_j}^\dagger(t) = \frac{1}{\sqrt{2}\;y_j} \left(
    y_j^2\,\hat\varphi_{\pmb{k}}{}_{ j} - i \,  
    \hat\pi_{\pmb{k}}{}_{  j}  \right)  \quad j=1,2 \, ,
\ee
or, in matrix representation 
\be
A_{\pmb{k}} (t)= U \, Y\,\hat  z_{\pmb{k}}(t) \, ;
\label{defa}
\ee
where
\be
\begin{split}
  &A_{\pmb{k}}(t) = \left(a_{{\pmb{k}}{}_1}(t), \, a_{{\pmb{k}}{}_2}(t),
  a_{{-\pmb{k}}{}_1}^\dagger(t), \, a_{{-\pmb{k}}{}_2}^\dagger(t) \,
\right)^t \, , \qquad Y =\begin{pmatrix} y & 0 \\
0 & y^{-1} \end{pmatrix} \, , \qquad y = \begin{pmatrix}
y_1 & 0 \\
0 & y_2 \end{pmatrix}  \, ,
\end{split}
\label{defy}
\ee
together with the obvious inverse relations.
The important physical difference is that we do not have a unique
option in the choice of the $y$ matrix; the choice of $N_{\pm}$ in the
definition of $\hat z_c$ (and that of $B$) was instrumental to get the Hamiltonian in the standard
form (\ref{hnorm})  for an harmonic oscillator.  
As before, the $a$-type
operators also define the associated vacuum state by
\be
a_{{\pmb{k}}{}_j} (t_0)|0_a \rangle =0  \, ,\qquad  \langle 0_a |0_a
\rangle =1 \, .
\ee
The Hamiltonian in the new basis is the following
\be
H_k =\ha \, A^\dagger_{\pmb{k}}(t) \, {\cal H}_A \,  A_{\pmb{k}}(t) \,
, \qquad  {\cal H}_A=U \, Y^{-1}{}^t \, {\cal H}_k \, Y^{-1} \,
U^\dagger \, .
\ee
As before $A_{\pmb{k}}(t)$ are in the Heisenberg picture and
\be
 A_{\pmb{k}}(t) = {\cal G}_A(t) \,  A_{\pmb{k}}(0) \equiv {\cal G}_A(t) \,
 A_{\pmb{k}} \, , \qquad {\cal G}_A(t)= U \, Y \, G(t) \, Y^{-1} \,
 U^\dagger \, ,
 \label{Aevolv}
 \ee
 and $G(t)$ gives the time evolution of $z_{\pmb{k}}$ according to
 (\ref{zevol}). By using (\ref{Aevolv}),  (\ref{zevol}), and that
 $G_c^\dagger \,  \,\Lambda_c^{(\pm)} \, G_c=\Lambda_c^{(\pm)}$, the
 Hamiltonian can be written in terms of $ A_{\pmb{k}}$ instead of
 $ A_{\pmb{k}}(t)$
\be
H_k =\ha \, A^\dagger_{\pmb{k}} \, {\cal H}_A \,  A_{\pmb{k}}\, ,
\ee
which shows that $H_k$ is time-independent. In general, $H_k$ can be
written in terms of ten real parameters which can be interpreted as
 squeezing and rotation parameters~\cite{Grain:2019vnq,Colas:2021llj} or equivalently as Bogolyubov coefficients.
 Explicitly we get
\be
{\cal H}_A = \begin{pmatrix}
  P & Q\\
  Q^\dagger & P
\end{pmatrix} \, 
%
\;\;{\rm where}\;\;
%
P= P^\dagger= \begin{pmatrix}
  F_1 & F_{12} \,e^{i\,\phi}\\
 F_{12} \,e^{-i\,\phi} & F_2
\end{pmatrix} \,, \quad
 Q=\begin{pmatrix}
R_1\,e^{i\,\Theta_1} & R_{12}\,e^{i\,\xi}\\
 R_{12}\,e^{i\,\xi} &  R_2\,e^{i\,\Theta_2}
\end{pmatrix} \, ;
 \ee
 with
 \be
F_1=\frac{d^2+m_1^2+y_1^4}{2 \, y_1^2} \, , \qquad F_2
=\frac{d^2+m_2^2+y_2^4}{2 \,  y_2^2} \, , \qquad F_{12} =\frac{d
  \left(y_1^2+y_2^2\right)}{4 \, y_1 \, y_2} \, , \quad \phi=\frac{\pi}{2}
\ee
\be
\begin{split}
&R_1 = \frac{d^2+m_1^2-y_1^4}{2  \,y_1^2} \, , \qquad R_2 =
\frac{d^2+m_2^2-y_2^4}{2  \,y_2^2} \, , \qquad R_{12} =\frac{d
  \left(y_1^2-y_2^2\right)  }{2 \,y_1 \, y_2} \, , 
  \\
&
\Theta_1=\Theta_2=0 \, , \; \xi = \frac{\pi}{2} \, .
\end{split}
\ee
Thus
\bea
H_k&=& \sum_{i=1}^2 \left[ F_i \, \left( a^\dagger_{{\pmb{k}}{}_i}\, a _{{\pmb{k}}{}_i}+
a^\dagger_{{-\pmb{k}}{}_i}\, a_{{-\pmb{k}}{}_i}+\text{h.c.} \right)+
R_i \, \left(e^{i\,\Theta_i} \, a^\dagger_{{\pmb{k}}{}_i}\,
  a^\dagger_{{-\pmb{k}}{}_i}+\text{h.c.} \right) \right]+\\
&&F_{12}\,e^{i\,\theta}\, \left ( a^\dagger_{{\pmb{k}}{}_1}\, a_{{\pmb{k}}{}_2}+
\ a^\dagger_{{-\pmb{k}}{}_1}\,  a_{{-\pmb{k}}{}_2} \right)+\text{h.c.} + R_{12}\;e^{i\,\xi} \left(
a^\dagger_{{\pmb{k}}{}_1}\, a^\dagger _{{-\pmb{k}}{}_2}+
 a^\dagger_{{\pmb{k}}{}_2}\, a^\dagger_{{-\pmb{k}}{}_1} \right)
+\text{h.c.} \, .
 \ea
The physical interpretation of the various terms is the following
\cite{Grain:2019vnq,Colas:2021llj}
\begin{itemize}
\item {\it Harmonic}: $F_{i=1,2}$ non-standard normalization of the
  number operator;
\item {\it Parametric}: $R_{i=1,2}$ gives rise to particle creation;
\item {\it Transferring}: $F_{12}$  transfers particles from one sector to the other;
\item {\it Entangling}: $R_{12}$ represents cross-sector particle creation.
\end{itemize}
By a suitable choice of $y_{1,2}$, one  can set $R_{12}=0$ and $R_1=0$ or
$R_2=0$.

Let us discuss the relation between the quantization performed by using the above explicit
covariant symplectic formalism and the more traditional one that makes an ansatz for quantum fields according to which they can be written,
in Fourier space, as linear combination of creation and annihilation
operators \cite{Grain:2019vnq}; for instance in the case of a single scalar field
\be
\hat\varphi_{\pmb{k}}(t)= \phi_k(t) \, b_{\pmb{k}}+  \phi_k(t)^\ast  \,
b_{\pmb{k}}^\dagger \; ,
\label{std}
\ee
where  $\phi_k(t)$ is a solution of the linear equation of motion in
Fourier space. The requirement that the field $\hat\varphi(t, \pmb{x})$
together with its conjugate momentum  $\hat\pi(t, \pmb{x})$ satisfy the
equal time canonical commutation rules gives a condition on the
Wronskian of the solutions $\phi_k(t)$ and $\phi_k(t)^\ast$ of the
equation of motion of the field. As matter of fact such condition is
equivalent to the symplectic character of the transformations relating
the field variables, the matrices $S$ and $N$ in our case. The
symplectic treatment is particular useful when the conjugate momenta
are not simply proportional to the time derivative of the fields; this
is the case in a gyroscopic system, see eq. (\ref{cjm}). To write the
quantum field in the form (\ref{std}) we can use (\ref{bdef}) and
(\ref{zt}) to get
\be
\begin{split}
&\hat z_{\pmb{k}}(t) = S \, N_\pm \, U^\dagger \, B_{\pmb{k}}(t) \, , \qquad
B_{\pmb{k}}(t)= U \, \hat z_{\pmb{k}}{}_c (t)= U \, G_{\pm}{}_c(t) \,
\hat z_{\pmb{k}}{}_c (0)= U \, G_{\pm}{}_c(t) \, U^\dagger \,
B_{\pmb{k}}(0) \\[.2cm]
& \qquad \Rightarrow \qquad \hat z_{\pmb{k}}(t) =  S \, N_\pm \,  G_{\pm}{}_c(t)
\, U^\dagger \,B_{\pmb{k}}(0) \equiv \mathbb{E}(t) \, B_{\pmb{k}}(0)  \,.
\end{split}
\ee
The time-dependent $4 \times 4$ symplectic matrix $\mathbb{E}(t)$ is determined by $U$,
the canonical transformations $S$ and $N_{\pm}$ and the symplectic
time evolution matrix $ G_{\pm}{}_c(t)$ and can be written in terms of
the suitable ``classical modes'' that can be read out from $\mathbb{E}(t)$;
in particular the expression for the quantum fields $\hat \varphi_{1,2}$ is the following 
\be
\begin{split}
  & \begin{pmatrix} \hat \varphi_{{\pmb{k}}{}_1}(t) \\
    \hat \varphi_{{\pmb{k}}{}_2}(t) \end{pmatrix} 
    =  L(t) \begin{pmatrix} b_1(\pmb{k})\\
  b_2(\pmb{k}) \end{pmatrix} + L^\ast(t) \begin{pmatrix} b_1(\pmb{k})^\dagger\\
  b_2(\pmb{k})^\dagger \end{pmatrix}  \, , \\[.2cm]
&
L(t)=\left(
\begin{array}{cc}
\mathbb{E}_{11}&\mathbb{E}_{12}\\
\mathbb{E}_{21}&\mathbb{E}_{22}
\end{array}
\right)=
\left(
\begin{array}{cc}
 \frac{\sqrt{T_{11}} \;e^{-i\,\omega _1\,t}}{2\,
   \sqrt{\omega _1} \,\sqrt{\omega _1^2-\omega _2^2}} &
   -\frac{2 \,i\, d \,\sqrt{\omega _2} \,e^{\mp \,i\,\omega _2\,t}}{\sqrt{\pm T_{22}}\, \sqrt{\omega
   _1^2-\omega _2^2}} \\
 -\frac{2\, i\, d \,\sqrt{\omega _1}\,\, e^{-i\,
   \omega _1\,t}}{\sqrt{T_{11}} \,\sqrt{\omega _1^2-\omega
   _2^2}} & \frac{\sqrt{\pm T_{22}}\, \,e^{\mp\,i \,\omega _2\,t}}{2\,
   \sqrt{\omega _2} \,\sqrt{\omega _1^2-\omega _2^2}} \\
\end{array}
\right) \, .
\end{split}
\label{modes}
\ee
Clearly the matrix $L$ and $L^\ast$ are just submatrices of
$\mathbb{E} $.
One can check that the quantum fields satisfy the  equations of motion (\ref{eqm}). 
Of course the same representation can be used for the quantum fields
expressed in terms of the alternative set of creation and
annihilation operators (\ref{defb}) and their modes
\be
 \hat z_{\pmb{k}}(t) =\tilde{\mathbb{E}}_{\pmb{k}}(t) \, A_{\pmb{k}}(0) \, ;
 \ee
by using (\ref{Aevolv}) and (\ref{defa}), the  symplectic matrix $\tilde{\bf
  E}_{\pmb{k}}(t)$ is given by
\be
\begin{split}
& \hat z_{\pmb{k}}(t)=Y^{-1}\;U^\dagger \, A_{\pmb{k}}(t)=Y^{-1}\;U^\dagger
\, {\cal G}_A(t) \, A_{\pmb{k}}(0) \\
& \qquad \Rightarrow
\tilde{\mathbb{E}}_{\pmb{k}} (t)=Y^{-1}\;U^\dagger \, {\cal G}_A(t) \, .
\end{split}
\ee
The classical modes in $\tilde{\mathbb{E}}_{\pmb{k}} (t)$ are
solutions of  the classical equations of motion, namely
$\dot{\tilde{\mathbb{E}}}_{\pmb{k}}=\Omega \, {\cal H} \,\tilde{\mathbb{E}}_{\pmb{k}}$.
It is interesting to note that  the initial conditions at $t=0$ can be
related to the freedom in the choice of the matrix $Y$, see \cite{Grain:2019vnq}.
The initial conditions on ${\cal G}_A(t) $ is related to the initial conditions
for the classical modes
\be
  {\cal G}_A(0)= \pmb{I} \, \quad\Rightarrow \quad \tilde{\mathbb{E}}_{\pmb{k}}(0)=Y^{-1}\;U^\dagger \,  
  =\left(
\begin{array}{cccc}
 \frac{y_1}{\sqrt{2}} & 0 & \frac{y_1}{\sqrt{2}} & 0
   \\
 0 & \frac{y_2}{\sqrt{2}} & 0 & \frac{y_2}{\sqrt{2}}
   \\
 -\frac{i}{\sqrt{2} y_1} & 0 & \frac{i}{\sqrt{2} y_1}
   & 0 \\
 0 & -\frac{i}{\sqrt{2} y_2} & 0 & \frac{i}{\sqrt{2}
   y_2} \\
\end{array}
\right)\ee
The relation among the classical modes $\mathbb{E}_{\pmb{k}}$ and
$\tilde{\mathbb{E}}_{\pmb{k}}$ and the corresponding sets of creation
and annihilation operators is a Bogolyubov transformation~\cite{Birrell:1982ix,Parker:2009uva,Mukhanov:2007zz,Grain:2019vnq}.

\section{Gyroscopic Systems and the Pais-Uhlenbeck Oscillator}

It is  intriguing to relate the anomalous region of stability of a gyroscopic system with
the Pais-Uhlenbeck oscillator~\cite{Pais:1950za}. 
The Pais-Uhlenbeck higher derivative Lagrangian can
be written as
\be
L_{\text{PU}}= \ha \left[ \ddot{\varphi}^2 - \left(\omega_1^2+
    \omega_2^2 \right) \dot{\varphi}^2+ \omega_1^2 \, \omega_2^2 \,
  \varphi^2 \right] \, .
\ee
that gives the following  fourth  order equation of motion
\be
\varphi^{(4)}+(\o_1^2+\o_2^2)\;\ddot\varphi+\o_1^2\;\o_2^2\;\varphi=0
\, .
\label{eqPU}
\ee
The solution is of the form $\varphi \sim\exp(i \, \omega \, t)$ and
$\omega$ satisfies exactly (\ref{eqomega}). Though a gyroscopic system
has at least 2 degrees of freedom, it is easy to see that (\ref{eqPU})
is equivalent to a system of second order coupled equations \cite{Banerjee:2013upa}
\be
\ddot\varphi_1+\mu_1\;\varphi_1-\rho_1\;\varphi_2=0,\qquad \qquad
\ddot\varphi_2+\mu_2\;\varphi_2-\rho_2\;\varphi_1=0
\ee
where the real constants $\mu_i$ and $\rho_i$  are constrained by
\be
\mu_1+\mu_2=\omega_1^2+\omega_2^2 \, , \qquad \mu_1\;\mu_2-\rho_1\, 
\rho_2=\omega_1^2 \, \omega_2^2 \, .
\qquad 
\ee
Exploiting the freedom in the choice of $\mu_i$ and $\rho_i$ it is easy to realize that the Pais-Uhlenbeck
oscillator admits different classically equivalent  Lagrangian
formulations~\cite{Banerjee:2013upa,Pavsic:2016ykq}. Depending on the choice of $\rho_i$,
the two second order equations  can be derived by two different Lagrangians of the
form
\bea
L_{\text{PU}_{a/b}}= \ha \left [\dot{\varphi_1}^2 \pm \dot{\varphi_2}^2-
  \frac{1}{2}\,\left(\mu_1 \, \varphi_1^2\pm \mu_2 \, \varphi_2^2 - 2\,\rho_1 \,
    \varphi_1 \, \varphi_2   \right) \right] \; \qquad {\rm
  for}\;\;\rho_1=\pm\rho_2 \, ;
\ea
After a Lagrangian field redefinition,
$L_{\text{PU}_a}$ leads to an Hamiltonian that is positive
defined\footnote{This is possible when $\omega_1^2 \neq
  \omega_2^2$.},  while $L_{\text{PU}_b}$ leads to an Hamiltonian that is not positive defined. 
Our representation of the
Pais-Uhlenbeck oscillator is rather different, as  it is evident form
the equations of motion (\ref{eqm})
\be
\ddot\varphi_1+m^2_1\;\varphi_1-2\;d\;\dot\varphi_2=0,\qquad
\ddot\varphi_2+m^2_2\;\varphi_2-2\;d\;\dot\varphi_1=0 \, ;
\ee
nevertheless they are still equivalent to (\ref{eqPU}), moreover the Hamiltonian $H$ is positive definite in the
region of normal stability (\ref{normal}) or  indefinite in the anomalous
region of  stability (\ref{anom}). It is interesting to realize that the equivalence at the
  level of equations of motion of the PU oscillator and
  $L_{\text{PU}_a}$,  $L_{\text{PU}_b}$  in general  is altered when
  interactions are introduced \cite{Pavsic:2013noa}.  For instance, if one
  introduces an interaction potential of the form  $\lambda
 \;\varphi_1^2\;\varphi_2^2$, only the Lagrangian $L_{\text{PU}_b}$
 generates equations of motion that are equivalent  to a PU oscillator
 with the same interaction. This is not the case for $L_{\text{PU}_a}$
 and for our gyroscopic system. However,  by introducing a
 non-dynamical field, one modify the Lagrangian to extend the
 equivalence also to other cases~\cite{Gross:2020tph}.

\section{Examples of Gyroscopic Systems}
 \label{bexp} 
In this section we give a number of specific examples of gyroscopic
system considered on a 
Friedmann-Robertson-Walker (FRW) cosmological
 background.  When
perturbations of the metric are considered, departing the homogeneous
FRW metric the gyroscopic nature of such a system is not altered but
the treatment is more involved; see~\cite{Celoria:2019oiu,Celoria:2020diz,Celoria:2021cxq}  for the discussion in the
context of inflation and the computation of primordial non-Gaussianity.
On general grounds in $1+3$ dimensions we can define out of $N$ scalar
fields the following set
of composite operators shift symmetric but in general not  invariant
under internal $SO(3)$
\be
C^{AB}=g^{\mu \nu} \, \de_\mu \Phi^A \, \de_\nu \Phi^B \, , \qquad
A,B=1,2 , \cdots, N \, . 
\ee
Depending  of what kind of vev the  fields develop  one can distinguish
the following cases giving a sketch of the operators involved.

\begin{itemize}
\item
All fields have time dependent   vevs and $C^{AB}$ is a singlet under
internal $SO(3)$ for any choice of $A,B$ and we have $N(N+1)/2$ operators. 
 
\item
  All fields have space-dependent  vev and to be consistent with the
  unbroken  $SO(3)_d$ diagonal group as discussed in section \ref{origin} the
  fields must be arranged  in $n$ triplets of  $SO(3)_d$, namely $\Phi^A \to
  \Phi_i^a$ with $a=1,2,3$ and $i=1,2, \cdots, n=N/3$. The basic combination  of fields 
  is
  \be
  B_{ij}^{ab} = g^{\mu \nu} \, \de_\mu \Phi^a_i \, \de_\nu \Phi^b_j \, , \qquad
a,b=1,2,3  \quad i,j=1,2, \cdot \cdots, N/3 \, 
\ee
from which one can form a number of  $SO(3)_d$ invariant operators
\be
\label{SSX}
X_{i_1,...,i_n}^{(S)}=\text{Tr}[B_{i_1 i_2}...B_{i_{n-1} i_n}] \, .
\ee
\item
 Finally, the most involved case is when the fields develop both space
 and time dependent  vevs. For simplicity let suppose that there is a
 single field $\Phi^0$ that has a time dependent vev and the $N-1=3\,n$
 remaining fields arranged as triplets: $\Phi_i^a$ with vev $\phi_i^a
 = x^a$. This time the basic building block conveniently  organized  according 
 to the $SO(3)_d$ transformations  are, besides (\ref{SSX}),
 \be
 \label{STX}
 X_{i_1 j_2,...,i_n}=\text{Tr} \left(Z_{i_1j_2} \cdots Z_{i_{n-1} i_n} \right)
 \, ,
 \ee
 where
 \be\label{STZ}
Z^{ab}_{ij} = (g^{\mu \nu} \, \de_\mu \Phi^0\, \de_\nu \Phi^a_i) \;( g^{\alpha \beta} \, \de_\alpha \Phi^0 \, \de_\beta \Phi^b_j)
\, .
\ee
\end{itemize}
The above operators showed  here does not exhaust the list of possible
single derivative
$SO(3)_d$ invariant operators. Indeed, many others can be built out of
$u_{ijk}^\mu =\epsilon^{\mu\nu\rho\sigma}\;\epsilon_{abc} \,
 \partial_\nu\Phi^a_i\,\partial_\rho\Phi^b_j\partial_\sigma\Phi^c_k
 $. Given the complexity of the most general case and
 to illustrate  the general picture  described in section \ref{origin}
  we consider the case of  two scalar
 fields $\Phi_1$ and $\Phi_2$ in a $1+1$ dimensional FRW 
 background\footnote{For simplicity we take the spatially
   flat  case.}. The generalization  to the case of 1+3 dimensions   is
 not very difficult. 
 In 1+1 dimensions a single scalar field can
 develop a spatially dependent vev still preserving  homogeneity and
 spatial translational invariance while to do same in $1+3$ dimensions in a
 rotational invariant way we need at least three dof. In 1+1
 we have just three basic operators.

Taking the metric
\be
g_{\mu\nu}=a^2\;\eta_{\mu\nu} \, ,
\label{FRW}
\ee
 the  most general  action is of the form
 \be
 S= \int d^2x \; \sqrt{- g} \; U(\de \Phi,\,\Phi )\, .
 \ee
 The number of operators is limited to
 \be\label{Xi}
 X_1 = g^{\mu \nu} \, \de_\mu \Phi^1\, \de_\nu \Phi^1 \, \qquad   X_2
 = g^{\mu \nu} \, \de_\mu \Phi^2\, \de_\nu \Phi^2 \, ,  \qquad X_{3} =
 g^{\mu \nu} \, \de_\mu \Phi^1\, \de_\nu \Phi^2 \, .   
 \ee
 \subsection{  Time dependent vevs} 
 Consider first the case where both scalars have a
 time-dependent vev, namely
\be
\Phi^1 =  \phi_1(t) + T_1 \, , \qquad \qquad \Phi^2 = \phi_2(t) + T_2 \, .
\ee
The Lagrangian $U$ has the form $U=U(X_1,X_2, X_3, \Phi_1, \Phi_2)$.
The kinetic matrix ${\cal K}$ has the following matrix elements
\be
\begin{split}
&  {\cal K} = \frac{1}{a^2} \begin{pmatrix} {\cal K}_{11} &  {\cal K}_{12} \\
     {\cal K}_{12} &  {\cal K}_{22} \end{pmatrix} \, ; \\  
& {\cal K}_{11}= 
 -2 \,a^2 \, U_{X_1}+4 \, \dot{\phi }_1^2  \,U_{X_1^2}+\dot{\phi
   }_2 \left[4 \, \dot{\phi }_1 \, U_{X_1 X_{3}}+\dot{\phi }_2 \,
     U_{X_{3}^2}\right] \, ;\\
   & {\cal K}_{12}=  -a^2 \,  U_{X_{3}} +2 \,
   \dot{\phi }_1^2 \,  U_{X_1 X_{3}}+\dot{\phi }_2 \,
   \left[\dot{\phi }_1  \,\left(4  \,U_{X_1
   X_2}+U_{X_{3}^2}\right)+2 \, \dot{\phi }_2 \, U_{X_2
   X_{3}}\right] \, ;\\
 & {\cal K}_{22}=-2  \,a^2 \, U_{X_2}+\dot{\phi }_1^2 \,
   U_{X_{3}^2}+4  \,\dot{\phi }_2 \, \left[\dot{\phi }_1 \, U_{X_2
   X_{3}}+\dot{\phi }_2 \, U_{X_2^2}\right] \, .
\end{split}
 \ee
\be
{\cal D} =\frac{1}{2}  \,\left[\dot{\phi }_1 \, \left(U_{\Phi _1 X_{3}}-2 \,
   U_{\Phi _2 X_1}\right)+\dot{\phi }_2 \, \left(2  \,U_{\Phi
   _1 X_2}-U_{\Phi _2 X_{3}}\right)\right]\;{ \cal J} \, .
\ee
\be
{\cal M}=-\left(
\begin{array}{cc}
 2\; U_{X_1} & U_{X_{3}} \\
 U_{X_{3}} & 2\; U_{X_2} \\
\end{array}
\right)\;k^2 +\left(
\begin{array}{cc}
 \alpha_{11} & \alpha_{12}  \\
 \alpha_{12}  & \alpha_{22}  \\
\end{array}
\right) \, ;
\ee
the explicit expressions for $\alpha_{ij}$ are omitted for sake of
brevity and will be not relevant for the discussion.
The main features are the following
\begin{itemize}
\item The kinetic matrix is $k$-independent and the off diagonal elements
  are related to the presence of the operators $X_{3}$ and $U_{X_1 X_2}\neq0$.
\item  In the presence of  shift symmetry: $U_{\Phi _i }=0$ and then  automatically ${\cal D}=0$.
\item The mass matrix   is  $k$-dependent.
\end{itemize}
As a physical example in 1+3 dimensions, we can consider two scalar fields fluids such that 
 ${\cal K}$ is diagonal (no kinetic
mixing) and with constant sound speeds;
\be
U(X_1,X_2,\Phi_1,\Phi_2) = A_1 \, X^{\frac{1+c_{s_1}^2}{2 \,  c_{s_1}^2}}_1 + A_2 \,
X^{\frac{1+c_{s_2}^2}{2 \,  c_{s_2}^2}}_2+ V(\Phi_1,\Phi_2) \, ;
\label{twof}
\ee
which at the background level gives $\dot\phi_{1,2}= a^{1-c_{s_{1,2}}^2} $
\be
{\cal K}= \begin{pmatrix} \frac{(1+c_{s_1}^2) a^{1+ c_{s_1}^2}}{c_{s_1}^2} & 0\\
  0 & \frac{(1+c_{s_2}^2) a^{1+ c_{s_2}^2}}{c_{s_2}^2}\end{pmatrix}\, ,
\qquad {\cal D}=0 \, , \qquad {\cal M} =  \begin{pmatrix} \frac{(1+c_{s_1}^2) a^{1+
    c_{s_1}^2}}{c_{s_1}^2} \, k^2 & a^4 \, m_{12}^2\\
  a^4 \, m_{12}^2  & \frac{(1+c_{s_2}^2) a^{1+ c_{s_2}^2}}{c_{s_2}^2} \,
  k^2 \end{pmatrix} \, .
\ee
The canonical fields can be introduced to get rid of ${\cal K}$ by
$\varphi={\cal K}^{-1/2} \varphi_c$; as a result the new quadratic Lagrangian
has ${\cal K}={\bf I}$, still ${\cal D}=0$ and
\be
M_c =\begin{pmatrix} c_{s_1}^2 \, k^2 -\frac{1}{4} (1+c_{s_1}^2)^2 {\cal
    H}^2 -\frac{1}{2} (1+c_{s_1}^2) {\cal
    H}' & \frac{a^{2- c_{s_1}^2-c_{s_2}^2} \, c_{s_1} \, c_{s_2} \, m_{12}^2}{ (1+c_{s_1}^2)^{1/2}  (1+c_{s_2}^2)^{1/2}}\\
 \frac{a^{2- c_{s_1}^2-c_{s_2}^2} \, c_{s_1} \, c_{s_2} \, m_{12}^2}{
   (1+c_{s_1}^2)^{1/2}  (1+c_{s_2}^2)^{1/2}} &  c_{s_2}^2 \, k^2 -\frac{1}{4} (1+c_{s_2}^2)^2 {\cal
    H}^2 -\frac{1}{2} (1+c_{s_2}^2) {\cal
    H}' \end{pmatrix} \, .
\ee
The mass matrix can be diagonalized by an orthogonal transformation
with a time dependent mixing angle $\theta$ which will inevitably lead to a
gyroscopic system with
\be
D  = \begin{pmatrix} 0 & -\dot{\theta} \\
 \dot{\theta} & 0 \end{pmatrix}  \, ; \qquad \text{tan(2} \, \theta) =
\frac{8 \, c_{s_1} \, c_{s_2}
  a^{3-c_{s_1}/2-c_{s_2}/2}}{(c_{s_1}^2-c_{s_2}^2)(1+c_{s_1}^2)^{1/2}(1+c_{s_2}^2)^{1/2}
  \left[ 2 \, {\cal H}' +(2+c_{s_1}^2 +c_{s_2}^2){\cal H}^2 \right]} \, . 
  \ee
Even the case of two non-canonical scalar fields with a mass mixing in a FRW
background is a gyroscopic system in disguise  \cite{Senatore:2010wk}.

\subsection{ Space-dependent vevs} 
\label{Space-dependent}
In this case it is convenient  to define as in four dimensions
$\Phi^i=x^i +\frac{\partial_x}{\sqrt{\vec \nabla^2}}\,S_i\;$ ($i=1,2$) and,
as discussed in section \ref{origin}, we need shift symmetry; thus the
generic Lagrangian has the  form $U(X_1, X_2,X_3)$ and for the
matrices ${\cal K}$, ${\cal D}$ and ${\cal M}$ we get 
\be
{\cal K}=-
\left(
\begin{array}{cc}
  2 \;U_{X_1}  &U_{X_{3}}  \\
 - U_{X_{3} } &  2\;
   U_{X_2}  \\
\end{array}
\right) \,  \qquad {\cal D}=0 \, ;
\ee
\be
\begin{split}
&  {\cal M} = -  k^2  \,\begin{pmatrix}  {\cal M} _{11} & {\cal M}_{12} \\
  {\cal M}_{12} & {\cal M}_{22} \end{pmatrix} \,  ; \\
& {\cal M} _{11}= \frac{4 \left(U_{X_1^2}+U_{X_1
      X_{3}}\right)+U_{X_{3}^2}}{a^2}+2 \, U_{X_1} \, ;\\
&  {\cal M}_{12}  =\frac{4 \, U_{X_1 X_2}+2 \, \left(U_{X_1 X_{3}}+U_{X_2
      X_{12}}\right)+U_{X_{3}^2}}{a^2}+U_{X_{3}} \, ;\\
&  {\cal M}_{22} =     \frac{4 \left(U_{X_2^2}+U_{X_2
   X_{3}}\right)+U_{X_{3}^2}}{a^2}+2  \,U_{X_2} \, .
\end{split}
\ee
In this case
\begin{itemize}
\item 
the off diagonal elements are induced by the presence of the operator $X_{3}$;
\item  we have always  ${\cal D}=0$;
\item the mass matrix is quadratic in $k$ and it is not diagonal when
  the operator $X_{3}$  is present and   $U_{X_1 X_2}\neq 0$.
\end{itemize}
An  explicit example in Minkowski spacetime can be
found in~\cite{Esposito_2020} where the effective field theory for the interactions between acoustic and gapped phonons was studied. In a FRW set up, we see that the non diagonally kinetic and mass matrices will induce an effective $D$ matrix once the Lagrangian will be rewritten in the canonical form (\ref{dc}).
 
 \subsection{  Mixed vevs} 
 \label{Mixed}
Finally, in the mixed vevs case we have 
\be
\Phi^1 =  \phi(t) + T \, , \qquad \qquad \Phi^2 =x+ \frac{\partial_x}{\sqrt{\vec \nabla^2}}\,S\, .
\ee 
Now the $SO(3)$ invariant operators are $X_1$ and $X_2$ of (\ref{Xi}) and 
\be
\tilde X_3= X_3^2
\,.
\ee
as can be  by deduced by (\ref{STX},\ref{STZ})
and the Lagrangian is then of the form $U(X_1,X_2,  \tilde X_3,\Phi_1)$. 
Omitting for simplicity the kinetic matrix, we have
\be
{\cal D}=\frac{ k  \;\left(2\; U_{X_1 X_2}-U_{\tilde X_{3}}\right)\;
   \dot\phi}{a^2}\; { \cal J} \, ;
\ee
\be
{\cal M}=-\left(
\begin{array}{cc}
  2\;k^2\; \left(a^2 U_{X_1}+U_{ \tilde{X}_{3}}\right)  & k \;\beta_{12} \\
 k \;\beta_{12}  &  2\;k^2\; \left(  U_{X_2}+2 \frac{U_{X_2^2}}{a^2}\right) +k^2\;\gamma\\
\end{array}
\right) +  \left(
\begin{array}{cc}
 \alpha_{11} & \alpha_{12} \\
 \alpha_{12} &  \alpha_{22} 
\end{array}
\right) \; ;
\ee
once again the expression  of $\alpha_{ij}$, $\beta_{12}$ and $\gamma$  are not relevant for the discussion.
The main features are
\begin{itemize}
\item the kinetic matrix is always diagonal; 
\item  The system is genuinely gyroscopic  being ${\cal D}\neq 0$ when
  $U_{\tilde {X}_{3}}\neq0 $ and  $U_{X_1 X_2}\neq0 $ with an overall  $k$ dependence.
\end{itemize}
The present case is the most interesting one and will be further studied in 1+3 dim
in the following.

\section{Bunch Davies vacuum}
\label{BDV}
The Bunch-Davies (BD)  vacuum  is the vacuum of election
to set the initial conditions for cosmological perturbations during a
de Sitter or quasi de Sitter period.
A simple and physical way to define the  BD vacuum is to invoke  the
equivalence principle according with  at very small scales gravity
does not influence local physics. 
 In this context, by choosing conformal time as in (\ref{FRW}), we impose that at early time, namely when $ a\to 0
 $, the gyroscopic system behaves as in a Minkowski space, namely the
 Lagrangian in such limit is  time independent. By taking 
$a=t^{2/(1+3\;w)}$, the required time-independent Lagrangian at early
  time is obtained when the matrices entering the gyroscopic system are of the
  form 
\ba
  \nonumber
{\cal K}=\begin{pmatrix}
 \bar\kappa_1\;{  a^{\xi_1}}  & 0 \\
 0 &  \bar\kappa_2\;{  a^{\xi_2}}\\
\end{pmatrix}\,,\quad 
d=\bar d\;{  a^{\varsigma+\frac{\xi_1+\xi_2}{2}}}\,,\quad
{\cal M}_{ij}=\bar m_{ij}+\hat m_{ij}\,{ a^{\eta_{ij}}} \, ;
\quad \dot\theta_k=0 \, ;
\ea
all quantities with a bar are constant in time and
\be
  \eta_{11}= \xi _1\leq 0,\;\;\eta_{22}=\xi _2\leq 0, \;\;\eta_{12}\geq \frac{\xi_1+\xi_2}{2}\, ,\;\;w<-\frac{1}{3}
  \, ,\;\; \varsigma\leq 0 \, .
  \ee
By using the procedure of appendix \ref{canform} to reach the canonical form (\ref{gyrcan})
for the matrices of a gyroscopic system, we get in the limit $a\to 0$
\be
\begin{split}
& {\cal L}^{(BD)}=\ha \, \dot{\varphi}^t \, \dot{\varphi} +d_c \,  \varphi^t 
\, {\cal J} \, \dot{\varphi} - \ha \, \varphi^t  \, M^{(BD)} \,
\varphi  \, ;\\[.2cm]
& d_c = \begin{cases} 0 & \varsigma \neq 0 \\
 \frac{\bar d}{\sqrt{\bar \k_1\,\bar \k_2}} &\varsigma =0
  \end{cases}
\, ;
\end{split}
\ee
the  constant mass matrix $M^{(BD)}$ is given by 
\be
M^{(BD)}=\begin{pmatrix}
 \frac{\hat{m}_{1,1}^2}{\bar{\kappa
   }_1}  &
   \frac{\hat{m}_{1,2}^2}{\sqrt{\bar{\kappa }_1} \sqrt{\bar{\kappa
   }_2}}\;\delta_{\eta_{12}}^{(\xi_1+\xi_2)/2} \\
  \frac{\hat{m}_{1,2}^2}{\sqrt{\bar{\kappa }_1} \sqrt{\bar{\kappa
   }_2}} \;\delta_{\eta_{12}}^{(\xi_1+\xi_2)/2}& \frac{\hat{m}_{2,2}^2}{\bar{\kappa
   }_2}  \\
\end{pmatrix}+
\delta_{\xi_i}^0\, \begin{pmatrix}
 \frac{ \bar{m}_{1,1}^2}{\bar{\kappa }_1} &
   \frac{ \bar{m}_{1,2}^2}{\sqrt{\bar{\kappa }_1}
   \sqrt{\bar{\kappa }_2}} \\\,
 \frac{  \bar{m}_{1,2}^2}{\sqrt{\bar{\kappa }_1}
   \sqrt{\bar{\kappa }_2}} & \frac{  \bar{m}_{2,2}^2}{\bar{\kappa
   }_2}  
\end{pmatrix} \, .
\ee 
Notice that the $d$ is different from zero only when $\varsigma=0$.
In a rotational invariant theory, the dependence of spatial momentum of
the mass terms is of the form 
\be
\hat m^2_{i,i}=c^2_{s_i}\;k^2\;\bar\kappa_i\, ;
\ee
and $c^2_{s_i}$ play the role of  non-trivial sound speeds.
At small scales for which $k$ is very large, ${\cal D}$ is relevant
only if $\bar d \propto k$. This is the case when fields which develop
both space and time dependent vev. An example  is supersolid
inflation~\cite{Celoria:2020diz,Celoria:2021cxq} where
$w=-1,\;\xi_1=-\xi_2=4,\;\eta_{12}=1,\;\varsigma=0$. 
On the contrary, when only fields which develop  time-dependent vev are
present, the matrix ${\cal D}$ is not important in the limit of large
$k$ used in the selection of the BD vacuum \cite{Senatore:2010wk}.
 
\section{Dynamical Dark Energy as a Gyroscopic System} 
\label{DDE}
One of the open questions in modern cosmology is the nature of dark
energy that is driving the present expansion of our universe. The
simplest option is to add a non-dynamical cosmological constant to the
Einstein equations. Alternatively one may try to device a dynamical
model for dark energy associated with a medium of some sort with
pressure $p$, energy density $\rho$ and an equation of state $p
\approx -\rho$.
 A perfect fluid with a single scalar degree of
freedom\footnote{Two additional degrees of freedom corresponding to
a single transverse vector are present, however, thank to the
conservation of the vorticity, their dynamics is trivial.} does not
work: the energy momentum conservation forces $\rho$ to be a constant
and one gets back to a cosmological constant. To move on one needs to
go beyond a perfect fluid and/or add degrees of freedom. As matter of
fact, four scalar fields $\{\Phi^A , \; A=0,1,2,3 \}$, three $\{\Phi^a
    , \; a=1,2,3 \}$ with an $\vec{x}$-dependent vev  and $\Phi^0$ with
    a time-dependent vev can be used to describe the  most general non-dissipative
 self-gravitating medium at the leading derivative expansion and the
 analysis of section (\ref{origin}) applies. In particular, a genuine
 ${\cal D}$ term is present and we are dealing with a gyroscopic
 system. The action is
\be
S= \, M_{pl}^2 \, \int d^4x \, \sqrt{-g} \, U(b, y, \chi, \tau_Y,
\tau_Z) \, ;
\label{ssol}
\ee
where
\be
\begin{split}
 &b=(\text{Det}[B^{ab}])^{1/2}\,  , \qquad
y= u^\mu \de_\mu \Phi^0 \, , \qquad 
\chi=(-g^{\mu\nu}\de_\mu\Phi^0\de_\nu\Phi^0)^{1/2} \, ,\\
& B^{ab}
= g^{\mu \nu} \, \de_\mu \Phi^a \, \de_\nu \Phi^b  \, , 
\qquad \tau_Y = \frac{\text{Tr}(B^2)}{\text{Tr}(B)^2} \, , \qquad \tau_Z
= \frac{\text{Tr}(B^3)}{\text{Tr}(B)^3}  \qquad a,b=1,2,3
\end{split}
\ee
and
\be
u^\mu = -\frac{\epsilon^{\mu \nu \alpha \beta} }{6 \, b \, \sqrt{- g}}
  \epsilon_{abc} \, \de_\nu \Phi^a \,  \de_\alpha \Phi^b \,  \de_\beta
  \Phi^c \, , \qquad u^2=-1 \, .
  \ee
  The energy-momentum tensor (EMT) has the form
  \be
  T_{\mu \nu} = (U- b \, U_b) g_{\mu \nu}  + \left(y \, U_y -b \, U_b \right) u_\mu \,
  u_\nu +\chi \, U_\chi \, v_\mu \,
  v_\nu  +Q_{\mu \nu}^{(Y)} \, U_{\tau_Y} +Q_{\mu \nu}^{(Z)} \,
  U_{\tau_Z}  \, ;
  \label{EMT}
  \ee
  with
  \be
  \qquad v_\mu =\chi^{-1} \, \de_\mu \Phi^0 \,  .
  \ee
  In flat space or on a spatially flat FRW spacetime\footnote{The bar
  denotes background quantities and in Minkowski spacetime $\bar b =
  \bar \chi = \bar y=1$.} $\bar{u}_\mu = \bar{v}_\mu$,
$Q_{\mu \nu}^{(Z)} =Q_{\mu \nu}^{(Y)} =0$ and the
EMT is the one of a perfect fluid with
\be
\bar \rho =-U +\bar \chi \, U_\chi  + \bar y \, U_y \, , \qquad \bar p=U -
\bar{b} \, U_b \, .
\ee
Depending on che choice of $U$, different equation of state for the
medium can be considered. In~\cite{Celoria:2017idi}  there were studied
models, dubbed $\Lambda$-media, featuring an exact equation of state
$p+\rho=0$ (i.e. $w=p/\rho=-1$), valid  not only at the background
level in a FRW metric but also at the  non-perturbative one; this is
the case by taking
\be
U(b,y,\chi, \tau_Y, \tau_Z) \equiv b^{1+w} \, U_{w}(b^{-w} \,  \chi ,
\;b^{-w} \, y,\;  \tau_Y, \;\tau_Z) \, .
\label{Ude}
\ee
To study stability, away from any possible Jeans instability, it is
sufficient to consider the limit of very large spatial momentum
$k$ and forget about the expansion of the universe and
  metric perturbations\footnote{As shown in \cite{Celoria:2017hfd} one gets the very same
    result by the full analysis of quadratic perturbation in perturbed
    FRW universe.}. The scalar fields fluctuate according with
  \be
  \Phi^0= \phi(t) + \pi_0 \, , \qquad \Phi^a= x^a + \pi^a \, .
  \ee
  The $\pi^a$ excitations are decomposed  according to $\pi^a=
\pi^a_\perp+\de_a \pi_L$. The transverse part $\pi^a_\perp$
with $\de_a \pi^a_\perp =0$ describes vector modes that are not
considered here while  $\pi_L$  represents 
phonon modes. As discussed in section \ref{origin}, the
  structure of the vevs of the scalar fields is such that the
  quadratic Lagrangian derived from (\ref{ssol}) has precisely the
  form (\ref{quadlc}) and reads
 \be
  {\cal L}_{ph}=\frac{ 
   \left(\bar p +\bar \rho +M_1 \right) }{2} \dot S ^2+M_0\,
  \dot T ^2+\frac{\left(M_1-2\, M_4\right) \;k}{2}\;
   \left( S\,\dot T - T\,\dot S \right)+
\left(M_3-M_2\right) \,k^2\,
   S^2+\frac{ M_1}{2}\, k^2\,
   T^2 \, ,
   \ee
   where we have set $S= k \, \pi_L$, $T=\pi_0$ and
   \be
   \begin{split}
&   M_0 = \ha \left(U_{\chi \chi} + 2 \, U_{y \chi}+ U_{yy} \right) \,
   , \quad M_1 =-U_\chi \, , \quad M_3 = \ha U_{bb} \, , \quad M_2 =
   \frac{U_{\tau_Y} + U_{\tau_Z}}{27} \, , \\
&M_4 = U_{b
  \chi}+ U_{by}-\ha U_\chi - U_y \, .
\end{split}
\ee
By defining the dimensionless associated parameters $\mathit{c}_i$ through $M_i=\bar \rho \,
\mathit{c}_i$ we have that  a $\Lambda$-medium with $w=-1$ gives the
constraints\footnote{This can be seen as the consequence of a  Lifshitz
  scaling symmetry~\cite{Endlich:2012pz}.}
\be
w=-1,\quad\mathit{c}_0=\mathit{c}_4,\quad
\mathit{c}_2=3\,\left(\mathit{c}_3-\mathit{c}_4\right) \, .
\ee
The positivity of the kinetic terms imposes
\be
\mathit{c}_1  >0,\quad \mathit{c}_0>0 \, .
\label{kpos}
\ee
In the {\it normal region} of stability (\ref{normal}) the positivity
of the mass matrix requires that
\be
\left(\mathit{c}_3-\mathit{c}_2\right)\leq0,\quad \mathit{c}_1 \leq0
\ee
which is clearly incompatible with (\ref{kpos}); thus there is no room
for stability in the normal region. Stability 
in the {\it anomalous region}  defined by (\ref{anom}) requires that
\bea
& \left(\mathit{c}_3-\mathit{c}_2\right)=3\,\mathit{c}_4-2\,\mathit{c}_3
\geq0,\quad \mathit{c}_1\geq 0 \, , \\[.2cm]
& 0< \mathit{c}_1\leq
   \mathit{c}_4-\sqrt{\mathit{c}_4
   \left(3 \,\mathit{c}_4-2 \,\mathit{c}_3\right)} \, ,
\ea
and no inconsistency is present and one easily check that  $U$ of the
form (\ref{Ude}) does the job.
A point worth to be stressed is that stability with $w=-1$  requires 
$\mathit{c}_2\neq0$ which signals the presence of a solid component in
the medium associated to the operators $\tau_Y$ and $\tau_Z$;
incidentally the very same operators turn on an anisotropic stress part
in the EMT (\ref{EMT}) and by the gravitational Higgs also generate a
mass for the graviton. Thus there is no fluid-superfluid medium that
is stable with $w=-1$. At the same time the actual realization of
stability also requires that $\mathit{c}_4\neq0$ together with
$\mathit{c}_1\neq0$ which means stability also needs for the presence  of
a superfluid component related to the operator $\chi$.
The bottom line is that one can model a dynamical dark energy with
$w=-1$ that is stable at the quadratic level, the price to be paid is
that the Hamiltonian is not positive definite and is connected with the
Pais-Uhlenbeck oscillator. The results in~\cite{Gross:2020tph}
indicates that even in the presence of non-linearities the existence
of unavoidable pathologies are far from being automatic.

\section{Conclusions}
\label{conc}
We analyzed the classical and quantum dynamics  of quadratic
non-dissipative gyroscopic system that  are characterized by a Lagrangian
with a term of the form $\varphi \, \dot{\varphi}$ that is non-trivial when at least two degrees
of freedom are present.
In Minkowski spacetime such a term naturally appears when one consider
a set of coupled scalar fields in which some fields  acquire a
space-dependent vev and others a time-dependent one, spontaneously breaking
the Lorentz group down to the rotation group $SO(3)$.
The  minimal number  of scalar degrees of freedom for a
 gyroscopic system is two and  they can be interpreted as  the Goldstone modes
 for the spontaneous breaking of temporal and spatial  translations
 but also as the phonon-like excitations 
  of a supersolid,  a medium which has a superfluid and a solid
 component. 
We studied the classical and quantum dynamics by
 using symplectic techniques. The system is classically stable in two
 different regions in parameter space. In what we call  {\it normal}
 region, the Hamiltonian is positive defined while in the {\it
   anomalous} region the Hamiltonian is not positive defined; indeed,
 after a suitable canonical transformation can be written as the sum of a
 standard harmonic oscillator and a ghost-like oscillator. As a result,
 a gyroscopic system in the anomalous region of stability is
 related to  the physics of the Pais-Uhlenbeck oscillator. In the
 anomalous region of stability a resonant behavior  in the 2-point
 correlation function can take place and it is intriguing that it is
 behind the slow instability found in~\cite{Gross:2020tph} when
 the modes of the standard and the ghost modes are coupled. The very
 same resonant behavior  is behind the maximization  of the entanglement
 when the ghost mode is traced over.
 \\
 On a time dependent background as FRW (we always retain rotational
 invariance), the definition of gyroscopic systems becomes ambiguous
 due to the possibility of performing a time-dependent field redefinition,
 however it is possible to identify a set of field redefinitions such
 that a generic Lagrangian with two scalar fields can be brought in the
 canonical form
 (\ref{gyrcan}) for which the kinetic matrix is the identity, the mass
 matrix is diagonal and there is a gyroscopic term that mixes the fields
 with their time derivative through the antisymmetric matrix $D$.
The presence in the canonical form of the Lagrangian of  a non zero
matrix $D$ is taken as the definition of a gyroscopic system. It turns
out that a non-trivial $D$ can be induced by a time-dependent non-diagonal
 kinetic and/or mass matrix in the original Lagrangian which can be important for the existence of the  Bunch-Davies vacuum.
 As  a result, in a time-dependent background,
two coupled fluids/superfluids and two   coupled solids can also be
gyroscopic,  see (\ref{twof}) and (\ref{Space-dependent}).
Finally, we have shown that dynamical dark energy  can be described as
a gyroscopic system in the anomalous region of stability.\\ \\
 {\bf ACKNOWLEDGMENTS}
\\
We thank Marco Celoria and Rocco Rollo for   participating to the
early discussions concerning the subjects analyzed  in this paper.
D.C.  is grateful for pleasant discussions with Thomas Colas
during the {\it Hot topics in Modern Cosmology}
Spontaneous Workshop XIV May, 2022, IESC Cargese, France.

 \appendix

\section{Canonical Form}
\label{canform}
A generic symmetric  time-dependent matrix ${\cal T}$  can be
diagonalized by an orthogonal transformation
\be
{\cal R}^t \, {\cal T} \, {\cal R}={\cal T}_d=
\begin{pmatrix} 
\tau_1 &0 \\
0 & \tau_2
\end{pmatrix} \,,\qquad 
{\cal R}=
\begin{pmatrix} \cos \theta &
  \sin \theta\\  -  \sin \theta&  \cos
  \theta \end{pmatrix} \, .
\label{rot}
 \ee
The kinetic matrix ${\cal K}$ in (\ref{quadlc}) is positive definite
with eigenvalues $\kappa_1> \kappa_2>0$ and ${\cal
  K}_d=\text{Diag}(\kappa_1 , \, \kappa_2)$. By using (\ref{rot}),
after the following field
 redefinition 
\be
\varphi=A_{\cal K}\;\varphi' ,\qquad A_{\cal K}={\cal R}_{\cal K} \,
{\cal K}_d^{-1/2} \, ,
\ee
one gets the Lagrangian
\be
{\cal L}'=\ha \, \dot{\varphi} ' \, {\cal K}' \, \dot{\varphi}'+ \varphi' 
\, {\cal D}' \, \dot{\varphi}' - \ha \, \varphi' \, {\cal M}' \, \varphi' \, ;
 \ee
where
\bea
K'&=&\pmb{I}, \\
  {\cal D}' &=&A^t\,D\,A-\dot \theta_{\cal K}\;\frac{\text{Tr}({\cal  K})}{\text{Det}({\cal K})^{1/2}}\;{\cal J}\;\equiv \;d'\,{\cal J},\qquad
\qquad d'=
\frac{d-\dot\theta_{\cal K} \, \text{Tr}({\cal
  K}) }{ \det({\cal K})^{1/2}} \, ;\\[.2cm]
{\cal M}'&=&A^t\, {\cal M}\,A-\dot{A}^t\, {\cal  K}\,\dot{A}+\frac{1}{2}\frac{d}{dt}\left(\dot{A}^t\, {\cal K}\, {A}+
  {A}^t\, {\cal K}\,\dot{A} \right) \, .
\ea
One can also diagonalize the mass with ${\cal M}'=\text{Diag}(\tilde m_1^2, \,
\tilde m_2^2)$ to arrive to the canonical form given in
(\ref{quadlc}) by a final field redefinition
\be
\varphi'={\cal R}_{\cal M} \,\varphi''=A_{\cal M}^{-1}\,\varphi \, .
\ee
The structure of the Lagrangian in the canonical form for a gyroscopic
system is given by
\be
{\cal L}''=\ha \, \dot{\varphi}'' \, \dot{\varphi}''+ \varphi'' \,
 D\, \dot{\varphi}'' - \ha \, \varphi'' \,  M\,
\varphi'' \, ;
\ee
where $\varphi''={\cal R}_{\cal M}^{-1}\, {\cal K}_d^{1/2}\,{\cal
  R}_{\cal K}^{-1}\,\varphi$ and
\bea
D &=& {\cal D}'-\dot\theta_{\cal K}\; {\cal J}=d_c\;{\cal
  J},\qquad 
  d_c=\frac{d}{  \text{Det}({\cal K})^{1/2}}-\dot\theta_{\cal
  K}\;\frac{\text{Tr}({\cal K})}{ \text{Det}({\cal K})^{1/2}}-2\,
\dot\theta_{\cal M}^2,\\
 M &=& {\cal M}'-2\,\dot\theta_{\cal M}^2\, \pmb{I}=
\begin{pmatrix}
\tilde{m}_1^2-2\,\dot\theta_{\cal M}^2 &0\\
0& \tilde{m}_2^2-2\,\dot\theta_{\cal M}^2\,
\end{pmatrix} \equiv  \begin{pmatrix} m_1^2 & 0 \\ 0 & m_2^2 \end{pmatrix}\, .
\ea

\section{Lagrangian Transformation}
\label{tdeptr}
When the matrices are in the canonical form (\ref{gyrcan}) but
time-dependent, the equations of motion derived from (\ref{quadlc})
are the following
  \be
  \ddot{\varphi} - 2 \, D  \dot{\varphi}
  + \left(M + \dot{D} \right) \varphi =0 \, .
  \label{geneq}
  \ee
Let us now show that is not possible by a Lagrangian field
redefinition to set $D=0$. Taking 
$\varphi = L \, \tilde \varphi$, we get the following new form for the
matrices in (\ref{quadlc}) 
\be
\begin{split}
& \pmb{I} \to K_n = L^t \, K \, L \, , \qquad D \to D_n = \dot{L}^t \, L
+ L^t \, D \, L \, , \\
&M \to M_n = L^t \, M \, L - \dot{L}^t 
\, \dot{L} - 2 \, L^t D \, \dot{L} \, .
\end{split}
\ee
To find $D_n=0$, $L$ has to satisfies
\be
\tilde D= L^t DL + \ha \left( \dot{L}^t L - L^t \dot{L} \right) =0
\quad 
\Rightarrow  \quad 2 \, D + L \dot{L}^t - \dot{L} L^t =0 \, .
\ee
Clearly there is no solution if $\dot{L}=0$. If $L$ is orthogonal
then the solution is 
\be
\dot{L} = D \, L \, .
\ee
However, by a time-dependent field redefinition of the form
\be
L =\begin{pmatrix} \cos \theta(t) &  \sin \theta(t)\\-  \sin \theta(t) &  \cos
  \theta(t)\end{pmatrix}  \, , \qquad \dot{\theta} =d \, ,
\ee
the resulting Hamiltonian takes the form
\be
H_n= \frac{1}{2} \, P^tP +\frac{1}{2} \, \tilde \varphi^t M_n(t) \tilde \varphi \,
\ee
and it is inevitably time-dependent.  The time dependence is rather
special and it is of the Floquet type, namely
\be
\tilde{H}(t) = H_n(t+T) \, , \qquad T= \frac{2 \pi}{d} \, .
\ee
Such time dependent Hamiltonian is often found in condensed matter
physics. In the canonical form $M$ is diagonal and thus we get
%
%
\be
M_n= \begin{pmatrix}
d^2 +m_1^2  \cos(d \, t)^2 + m_2^2  \sin(d \, t)^2 &
\left(m_1^2 -m_2^2 \right) \sin(d \, t) \, \cos(d\, t) \\
\left(m_1^2 -m_2^2 \right) \sin(d \, t) \, \cos(d \, t) &
  d^2 +m_2^2  \cos(d\, t)^2 + m_1^2  \sin(d \, t)^2
\end{pmatrix} \, .
\ee
%
\section{Hamiltonian Diagonalization}
\label{hamdia}

The Hamiltonian diagonalization requires a symplectic
decomposition (congruence transformation). According  
to the Williamson theorem \cite{Nicacio:2021zmc}, if ${\cal H} $ is a {\it real symmetric
  positive} matrix of order $2\,n$, there exists a symplectic matrix
$S$ such that \ref{wil} holds and given $\Lambda_{\cal
  H}=\text{diag} \left(\lambda_j =i\,\omega_j\right)$ we have
$\det \left(\Omega\,{\cal H}\pm i\,\omega_j\,{\bf I}_{4\times 4} \right)=0$ and the matrix $S$ admits the decomposition
\be
S=\sqrt{\Lambda_{\cal H}}\,{\bf O}\,\sqrt{{\cal H}^{-1}} \, ;
\ee
where ${\bf O}$ is an orthogonal matrix satisfying
\be
{\bf O}\,\sqrt{{\cal H}}\,\Omega \,\sqrt{{\cal H}}\,{\bf
  O}^t=\Lambda_{\cal H}\,\Omega \, .
\ee
All the eigenvalues of the matrix $\Omega\, {\cal H}$ are purely
imaginary and
\be\label{wil}
S^t\cdot \Omega\cdot {\cal H}\cdot S= \text{diag} \left( i \, \omega_1,
  \, - i \, \omega_1, \,  i \, \omega_2,
  \, - i \, \omega_2 \right) \, .
\ee
Classical Stability requires that the fundamental frequencies
$\Omega_i\in{\bf C}$ and $\lambda_i^*=-\lambda_i$.
Once the Hamiltonian is diagonal, the Hamilton equations (\ref{Heq})
are simple to solve.
Imposing the initial conditions $ z_{\pmb k}(t =0)=z_0$, we can
formally write the solution of the above system as
\be  z_{\vec k}(t)=G_{k} (t,\,t_0)\cdot z_0 \, ;
\ee
where  $G_k(t,\,t_0)$ satisfies 
\be
\partial_t G_{ k} (t)=\Omega\cdot {\cal H}_{k} \cdot
G_{k} (t,\,t_0),\qquad G_{k} (t_0,\,t_0)={\bf I}_{4\times 4} \, .
\ee
The solution is
\be
G_{k}(t,\,t_0)=T_{\tau}\;e^{\int^t_{0}d\tau\;\Omega\cdot {\cal H}_{k}
  (\tau)} \, .
\ee
When ${\cal H}_{k} $ is time independent the $T$ ordering disappears  and we have
\be
G_{k} (t)= e^{\Omega\cdot {\cal H}_{k} \;t} \, .
\ee
The time evolution matrix $G$ is symplectic:
\be
\{z,\,z^\dagger\}=G\cdot \{z_0,\,z^\dagger_0\}\cdot
G^\dagger\;\;\Rightarrow\;\;\Omega=G\cdot \Omega\cdot G^\dagger=G\cdot
\Omega\cdot G^t,\, ;
\ee
with a real $G$.

\bibliographystyle{unsrt}  
\bibliography{biblio}

\end{document}